\documentclass[fleqn,12pt,twoside]{article}
\usepackage{espcrc1}

\usepackage{amssymb,amsmath}
\usepackage{caption}
\usepackage{graphicx}
\usepackage{mathrsfs}
\usepackage{dcolumn}
\usepackage{booktabs}
\usepackage{multirow}
\usepackage{relsize}
\usepackage{exscale}
\usepackage{subfigure}
\usepackage{soul}

\newcommand{\nuc}[2]{\ensuremath{^{\text{#1}}\text{#2}}}

\newcommand{\difd}{\mathrm{d}}

\newcommand{\fig}[1]{\mbox{Figure~\ref{#1}}}
\newcommand{\tab}[1]{\mbox{Table~\ref{#1}}}
\newcommand{\sect}[1]{\mbox{Section~\ref{#1}}}
\newcommand{\equ}[1]{Eq.~(\ref{#1})}

\title{Two-neutron overlap functions for \nuc{6}{He} from a microscopic structure model}
\author{I.~Brida\address[NSCL]
                        {NSCL and Department of Physics and Astronomy,
                         Michigan State University, \\ 
                         East Lansing, MI 48824, USA}
                \address[ANL]
                        {Physics Division, Argonne National 
                        Laboratory, Argonne, IL 60439, USA}
                \thanks{Corresponding author: brida@anl.gov},
        F.M.~Nunes\addressmark[NSCL]
}

\begin{document}
\maketitle

\begin{abstract}
A fully antisymmetrized microscopic model is developed for light two-neutron halo nuclei using a hyper-spherical basis to describe halo regions. 
The many-body wavefunction is optimized variationally. The model is applied to \nuc{6}{He} bound by semi realistic Minnesota nucleon-nucleon forces. The two-neutron separation energy and the radius of the halo are reproduced in agreement with experiment. Antisymmetrization effects between \nuc{4}{He} and halo neutrons are found to be crucial for binding of \nuc{6}{He}. We also properly extract two-neutron overlap functions and find that there is a significant increase of 30\%-70\% in their normalization due to microscopic effects as compared to the results of three-body models.
\end{abstract}


\section{Introduction}
\label{sect intro}

In some light nuclei, the proximity to particle emission thresholds allows loosely bound nucleons to tunnel out into the classically forbidden region and form what is typically referred to as a {\it halo}. One neutron halos, such as \nuc{11}{Be} and \nuc{19}{C}, and two-neutron halos, such as \nuc{6}{He} and \nuc{11}{Li}, are the most common, although there are nuclear states with the halo formed by protons or more than two neutrons. All these systems have very small separation energies and unusually large matter radii, when compared to neighboring nuclei, and many of their properties are determined by the long-range part of the many-body wavefunction. Halo systems are not specific to nuclear physics; a review covering a number of fields can be found in \cite{jensen04}. In this work, we focus on two-neutron halo nuclei in which long-range Coulomb effects are not present.

Traditionally, the theory of halo nuclei has been dominated by simple few-body models. These models place their faith on the fact that the valence particles forming the halo are partially decoupled from the rest of the system, the core. In this frame of mind, one typically assumes that the core degrees of freedom are completely frozen, while meticulously treating the relative motion between the valence nucleons and the core. Then, the many-body problem containing a two-neutron halo reduces to a core+n+n three-body one. For an overview of three-body techniques applied to halo nuclei, see for example \cite{zhukov93}.

The crucial assumption of few-body models, a macroscopic inert core, is a mixed blessing: on one hand, it allows one to focus on the long-range core-valence correlations, on the other hand, it is undoubtedly a crude simplification of the many-body problem. Stemming from the inert core employed, the main drawbacks of three-body models are: i) unknown properties of the core, i.e. the internal details of the core are left out completely and, only when needed, they are provided in an ad hoc manner, ii) the improper treatment of antisymmetrization \cite{thompson00}, iii) the use of effective core-nucleon interactions constructed case by case in an ambiguous manner, and iv) the fact that even when the two-body interactions are fitted to (some) properties of all two-body subsystems, the resulting three-body separation energy is (much) smaller than the experimental value \cite{zhukov93,funada94}. This pathological lack of binding is often cured with an empirical three-body force or the two-body forces are refitted to deliver the right three-body binding. Furthermore, there are indications that for reaction calculations three-body wavefunctions may require additional renormalization to account for microscopic effects missing in the inert-core approximation \cite{timofeyuk01}. Some improvement over the inert-core picture is provided by few-body models incorporating collective excitations into the core \cite{vinh95,nunes96,brida06}; they, however, still suffer from the above-mentioned drawbacks of few-body models, and in addition are limited to applications where the core is a good rotor or vibrator. Given all these drawbacks, the predictive power of three-body models of two-neutron halo nuclei is rather limited. 

The above-mentioned drawbacks of few-body models are eliminated in microscopic (cluster) models of light nuclei. Over the last decade, there have been tremendous advances in brute force ab-initio methods and due to increasing computational power, these models have improved their accuracy and predictability. Amongst these, the Green's function Monte Carlo (GFMC) \cite{pieper01,pieper05}, the no-core shell model \cite{navratil09,navratil98,navratil01}, and molecular dynamics models \cite{feldmeier00,itagaki03,neff05} have been successfully applied to light s- and p-shell nuclei. Somewhere between few-body and ab-initio models are microscopic cluster models in which some degrees of freedom are frozen to reduce the complexity of the many-body problem. Of particular relevance to this work is the stochastic variational method (SVM) \cite{varga95,svm-book} which inspired several aspects of the model here presented. Applications of SVM and some other cluster models to two-neutron halo nuclei can be found for example in \cite{arai99,varga02,korennov04,armstrong07,vasilevsky01}.

Despite the success of existing microscopic (cluster) models applied to non-halo nuclei and their ability to reproduce basic properties of some halo species, such as (three-body) binding energies and radii, questions may arise about their ability to capture the long-range halo correlations. These correlations are important because the halo nucleons spend a considerable amount of time in regions distant from the core and, generally, it is this part of configuration space that contributes the most to reaction observables. As is known from three-body models, the wavefunction describing a two-neutron halo nucleus ought to fall off exponentially. Existing microscopic structure models do not pay close attention (if any) to asymptotic regions, instead they use the binding energy to assess the convergence towards an eigenstate. The convergence of the energy, however, does not necessarily guarantee the convergence of the wavefunction in long-range regions. Moreover, most microscopic models exploit computationally tractable bases, most commonly Gaussians of one sort or the other. In principle, it should be possible to capture the slower exponential decay by using a large Gaussian basis, but as argued in \cite{jensen04}, quality precedes quantity when it comes to halo nuclei; that is the shape of the basis functions matters more than the size of the basis.

In addition, existing microscopic structure models fail to provide input to reaction calculations of two-neutron halo nuclei. To feed reaction calculations, by themselves formulated in a few-body picture, one would have to extract information about halo particles from a full microscopic wavefunction which is a non-trivial task. Even though recently we have witnessed some progress in this direction for two-body non-halo projectiles \cite{nollett01,navratil04}, most microscopic structure theories are still far from providing such few-body-like information about three-body-like halo nuclei. It is for this reason that in reaction calculations the structure of halo nuclei is taken from few-body models despite all their drawbacks.

It is obvious that both few-body and microscopic structure models have their appealing aspects as well as drawbacks. It is the aim of this work to combine the best of the two worlds: to develop a fully microscopic model for light two-neutron halo nuclei bound by nucleon-nucleon interactions that would
describe simultaneously short- and especially long-distance regions. The novelty of our model is the integration of few-body and ab-initio methods. We use a basis that describes short-range correlations and at the same time preserves the correct three-body asymptotics. This is achieved at a high computational cost since matrix elements cannot be evaluated analytically. Our ultimate goal is to provide microscopic structure information to be used in reaction calculations involving two-neutron halo nuclei.

In this paper our model is applied to the simplest two-neutron halo nucleus, \nuc{6}{He}. Preliminary results for this case have been published in \cite{brida08}. Since neither \nuc{5}{He} thought of as \mbox{\nuc{4}{He}+n} nor the \mbox{n+n} subsystem are bound, \nuc{6}{He} is a Borromean system \cite{zhukov93}. To provide input for reaction calculations involving this nucleus, two-neutron overlap functions are properly computed from our microscopic model and, to our best knowledge, they are for the first time expressed in hyper-spherical coordinates. By doing so, these functions are directly applicable to some reaction calculations and they can be compared directly to three-body wavefunctions.



This paper is organized as follows. In \sect{sect theory}, the theoretical formulation of our model is developed. \sect{sect computation} outlines some technical details and numerical procedures employed. Results for \nuc{6}{He} are 
discussed in \sect{sect results}. A summary can be found in \sect{sect outlook} along with an outlook of possible future developments.


\section{Theory}
\label{sect theory}

The model here presented has as its primary goal the applicability to reactions involving light two-neutron halo nuclei and thus needs to capture few-body long-distance features of these systems. A two-neutron halo nucleus will be described by an antisymmetrized product of a microscopic core and a valence part consisting of two neutrons, or schematically \mbox{$\Psi = \mathcal{A}^{core-val} (core \times valence)$}. The terms ``core" and ``valence" refer to distinct pieces of the wavefunction prior to the action of the core-valence antisymmetrizer $\mathcal{A}^{core-val}$ upon which nucleons from the two parts of the wavefunction become indistinguishable.
At large distances, our wavefunction decouples into the three-body-like form \mbox{$\Psi \longrightarrow core\times n\times n$} of a desired shape most naturally expressed in Jacobi coordinates. It was therefore our choice from the early start to express the entire wavefunction in Jacobi coordinates. As an additional benefit of using these coordinates, translational invariance is guaranteed. All details of the model can be found in \cite{brida-thesis}.

More precisely, the microscopic core with a fixed total angular momentum and parity $J_{core}^{\pi}$, and isospin $T_{core}$ with projection $M_{T_{core}}$ is described by an antisymmetrized wavefunction $\Phi_{ J_{core}^{\pi} \, T_{core} \, M_{T_{core}} }$ corresponding to the core's ground state. If desired, excited states of the core can be included in the future. The valence terms are drawn from a suitable hyper-spherical basis. Each valence basis function $\psi$ carries total angular momentum $J_{val}^{\pi}$, and isospin $T_{val}$ with projection $M_{T_{val}}$, which coupled to the core's quantum numbers give $J^{\pi}$, $T$, and $M_T$ for the full system. For neutron-rich two-neutron halo nuclei, such as \nuc{6}{He} and \nuc{11}{Li}, $T_{val}=1$ and $M_{T_{val}}=-1$, and the core-valence isospin coupling is trivial. It is understood that parities $\pi$ carry the same subscripts as their corresponding $J$ and that $\pi=\pi_{core} \pi_{val}$. Then, the total wavefunction is written as:
\begin{equation}\label{eq MiCH wavefunction}
   \Psi_{J^{\pi} M_J \, T \, M_T} =
   \sum
   c_{J_{core}^\pi \, \Gamma_{val} \, J_{val}^{\pi}} \,
   \mathcal{A}^{core-val}
   \left[
      \Phi_{ J_{core}^{\pi} \, T_{core} \, M_{T_{core}} }
      \otimes
      \mathcal{A}^{val} \,
      \psi_{\Gamma_{val} \, J_{val}^{\pi} \, 1-1}
   \right]_{J^{\pi} M_J \, T \, M_T}.
\end{equation}
By acting on valence particles only, the operator \mbox{$\mathcal{A}^{val}=\sum_1^2 (-1)^p P$} antisymmetrizes the valence part, whereas the core-valence antisymmetrizer \mbox{$\mathcal{A}^{core-val}=\sum_1^{A(A-1)/2}(-1)^p P$} permutes valence
particles with those inside the core. In these operators, $P$ are permutation operators, $p$ are permutation parities, and $A=A_{core}+2$ is the mass number of the halo nucleus. The meaning of $\Gamma_{val}$ is explained in \sect{sect valence}.

The wavefunction in \equ{eq MiCH wavefunction} is constructed in two steps. First, the wavefunction of the core as a free nucleus is built within SVM, as briefly described in \sect{sect core}. Unlike many microscopic cluster models employing a simple $0s$-harmonic oscillator approximation to \nuc{4}{He}, we use the best possible wavefunction for the core obtainable within SVM. By doing so, we hope to attenuate the problem of underbinding relative to the three-body threshold \cite{arai99,csoto93}. Once the core wavefunction is optimized, its parameters remain unchanged in the subsequent minimization procedure. This implies that, as in other microscopic cluster models, distortion of the core due to valence neutrons is not accounted for explicitly, although some distortion is delivered through the core-valence antisymmetrization. In the second step, the valence part is constructed by drawing terms $\psi_{\Gamma_{val} \, J_{val}^{\pi} \, 1-1}$ from a hyper-spherical basis described in \sect{sect valence}. Valence basis functions contain discrete as well as non-linear continuous parameters that are optimized variationally, along with the linear coefficients $c$ in \equ{eq MiCH wavefunction}, by minimizing the expectation value of the Hamiltonian $H$:
\begin{equation}\label{eq expectation value of energy}
   E = \frac{ \langle \Psi | H | \Psi \rangle }
            { \langle \Psi |     \Psi \rangle }
\end{equation}
as outlined in \sect{sect computation}. To feed reaction calculations involving nuclei of interest, two-neutron overlap functions are extracted from the resulting optimized wavefunctions, as presented in \sect{sect overlap}.


\subsection{The core in SVM}
\label{sect core}

Our choice of the microscopic model for the core has been motivated by the following factors. The model should provide accurate structure for the core; it needs to be extendable to cores heavier than \nuc{4}{He}; and it must handle central and non-central nucleon-nucleon forces. Unlike for the valence part, there is no need to impose special asymptotic requirements on the core. Finally, as mentioned earlier in this section, the wavefunction of the core should be expressed in Jacobi coordinates. It is for these reasons that the SVM model seemed to be the most appropriate. In this section, basic ingredients of SVM applied to \nuc{4}{He} are outlined; more details on the method can be found in \cite{varga95,svm-book,brida-thesis}.

\begin{figure}[t!]
   \begin{center}
      \includegraphics[width=0.5\textwidth]
                      {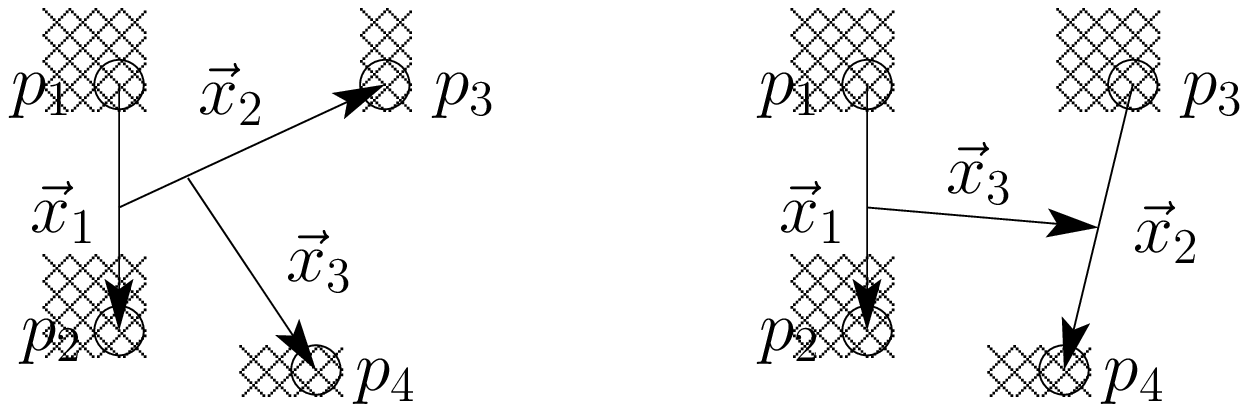}
      \caption{Sets of relative Jacobi coordinates $\vec{x}$ for a four-particle                ($p_1,\ldots,p_4$) system: K-like (left) and H-like (right). Each
               Jacobi coordinate connects centers of masses of subgroups of
               particles.}
      \label{fig core coordinates}
   \end{center}
\end{figure}

In SVM, the core wavefunction is written as a linear combination of basis functions $\phi$:
\begin{equation}\label{eq SVM wavefunction}
   \Phi_{J_{core}^{\pi} \, T_{core} \, M_{T_{core}}} =
      \mathcal{A}^{core}
      \sum
      c_{\Gamma_{core} \, J_{core}^{\pi} \, T_{core} \, M_{T_{core}}}
      \phi_{\Gamma_{core} \, J_{core}^{\pi} \, T_{core} \, M_{T_{core}}}
\end{equation}
with the projection of $J_{core}$ suppressed. The operator $\mathcal{A}^{core}$ antisymmetrizes particles inside the core. Basis states $\phi$ take the form of correlated Gaussians expressed in Jacobi coordinates. For a four-particle system, there exist two
different---K- and H-like---sets of Jacobi coordinates $\vec{x} \equiv \{ \vec{x}_1, \vec{x}_2, \vec{x}_3\}$ as shown in \fig{fig core coordinates}. To accelerate the optimization of the core's wavefunction, both Jacobi sets enter \equ{eq SVM wavefunction} as they invoke different inter-particle correlations. For \nuc{4}{He}, a basis term in either Jacobi basis is given by:
\begin{equation}\label{eq SVM single gaussian}
   \phi_{\Gamma_{core} \, J_{core}^{\pi} \, T_{core} \, M_{T_{core}}}
   \left(\vec{x},{\bf A}\right) =
   \exp \left( -\frac{1}{2}x{\bf A}x \right)
   \left[
   \theta_{l_1 l_2 l_3 L_{12} L}(\vec{x})
   \otimes
   \chi_{S_{12} S_{123} S}
   \right]_{J_{core}^{\pi}}
   \tau_{T_{12} T_{123} T_{core} M_{T_{core}}}.
\end{equation}
The function $\theta_{l_1l_2l_3L_{12}L}(\vec{x})$ is taken as a vector-coupled product of solid harmonics \cite{varga06}:
\begin{equation}\label{eq core orbital part}
   \theta_{l_1l_2l_3L_{12}L}(\vec{x}) =
   \left[
      \left[ \mathcal{Y}_{l_1} (\vec{x}_1)
             \otimes
             \mathcal{Y}_{l_2} (\vec{x}_2)
      \right]_{L_{12}}
      \otimes
      \mathcal{Y}_{l_3} (\vec{x}_3)
   \right]_L.
\end{equation}
The spin $\chi_{S_{12}S_{123}S}$ and isospin $\tau_{T_{12}T_{123}T_{core}M_{T_{core}}}$ parts consist of successively coupled single-particle spins and isospins:
\begin{eqnarray}
   \chi_{S_{12}S_{123}S} & = &
   \left[
      \left[
         \left[
            \chi_{p_1} \otimes \chi_{p_2}
         \right]_{S_{12}}
         \otimes \chi_{p_3}
      \right]_{S_{123}}
      \otimes \chi_{p_4}
   \right]_S,
   \\
   \tau_{T_{12}T_{123}T_{core}M_{T_{core}}} & = &
   \left[
      \left[
         \left[
            \tau_{p_1} \otimes \tau_{p_2}
         \right]_{T_{12}}
         \otimes \tau_{p_3}
      \right]_{T_{123}}
      \otimes \tau_{p_4}
   \right]_{T_{core}M_{T_{core}}}.
\end{eqnarray}
%

For \nuc{4}{He}, the Gaussian part in \equ{eq SVM single gaussian} contains a \mbox{$(3 \times 3)$}-dimensional positive-definite, symmetric matrix ${\bf A}$, specific to each basis term. The quadratic form \mbox{$x{\bf{A}}x$} involves scalar products of Jacobi vectors:
\begin{equation}\label{eq exponent in SVM}
   x {\bf A} x \equiv \sum_{i,j=1}^3
      {\bf A}_{ij}\vec{x}_i \cdot \vec{x}_j.
\end{equation}
Due to the symmetry requirement, the matrix ${\bf A}$ has only 6 independent elements and they are considered non-linear continuous variational parameters. Note that, although the Gaussian in \equ{eq SVM single gaussian} as a whole is a spherically symmetric object, it still carries angular information due to cross terms $\vec{x}_i\cdot\vec{x}_j$ when the matrix ${\bf A}$ is non-diagonal as considered here. In such a case, numbers $l_1,l_2,l_3$ in \equ{eq core orbital part} loose their meaning of orbital momentum quantum numbers and can be treated as discrete variational parameters, instead.

The composite index $\Gamma_{core}$ comprises other (quantum) numbers, elements of the matrix ${\bf A}$, and the Jacobi channel identifier K or H, i.e. $\Gamma_{core}=$\{$l_1, l_2, l_3$, $L_{12}$, $L$, $S_{12}$, $S_{123}$, $S, T_{12}, T_{123}$, K/H, {\bf A}\}. The sum in \equ{eq SVM wavefunction} was left without a summation index because in SVM the wavefunction $\Phi$ is constructed term by term by minimizing the expectation value of energy with all components of $\Gamma_{core}$ optimized stochastically. Linear expansion coefficients $c$ in \equ{eq SVM wavefunction} are determined via energy matrix diagonalization.


\subsection{The valence part in the hyper-spherical formalism}
\label{sect valence}

To inspire the form of valence functions in \equ{eq MiCH wavefunction}, let us first consider a few-body approach in which a two-neutron halo nucleus is treated as a three-body core+n$_1$+n$_2$ system. From among all three-body models, we adopt the formalism of \cite{thompson04} and \cite[pg. 278]{nunes-book}. For all necessary details, see \cite{brida-thesis}.

\begin{figure}[t!]
   \begin{center}
      \includegraphics[width=0.55\textwidth]
                      {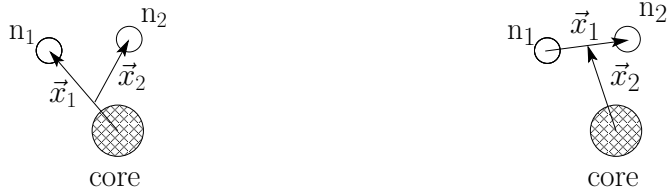}
      \caption{Sets of relative Jacobi coordinates for a three-body 
               core+n$_1$+n$_2$ system: Y-like (left) and T-like (right). Each 
               Jacobi coordinate connects centers of masses of subgroups of 
               objects. These coordinates are not to be confused 
               with those for the core in \fig{fig core coordinates}.}
      \label{fig valence coordinates}
   \end{center}
\end{figure}

For a three-body core+n$_1$+n$_2$ system, one can define two different---Y- and T-like---sets of Jacobi coordinates $\vec{x} \equiv \{\vec{x}_1, \vec{x}_2\}$ as shown in \fig{fig valence coordinates}. These coordinates are further transformed into hyper-spherical coordinates \{$\rho,\theta,\hat{x}$\} where $\hat{x}\equiv \{\hat{x}_1, \hat{x}_1\}$ are spherical angles corresponding to vectors $\vec{x}$. The hyper-radius $\rho$ is related to the overall size of the system, while the hyper-angle $\theta$ contains information about relative magnitudes of $\vec{x}_1$ and $\vec{x}_2$. In three-body models, the use of hyper-spherical coordinates is motivated by the facts that: i) the three-body Schr\"odinger equation reduces to a one-dimensional hyper-radial differential equation, and ii) more importantly for us, in the absence of long-range forces the three-body wavefunction of Borromean systems decays asymptotically for $\rho \rightarrow \infty$ as:
\begin{equation}\label{eq 3body asymptotics}
   \rho^{-5/2}\exp(-\kappa \rho), \qquad \kappa^2 = 2m\left|E_{3body}\right| / \hbar^2,
\end{equation}
where $E_{3body}$ is the binding energy relative to the three-body core+n+n threshold, and $m$ is the mass of a nucleon. In hyper-spherical coordinates, the volume element becomes:
\begin{equation}\label{eq hyper-volume}
   \difd V =
   \left( \mu_1\mu_2 \right)^{-3/2}
   \rho^5 \sin^2 \theta \cos^2 \theta \, \difd \rho \, \difd \theta \,
   \difd\hat{x}_1 \, \difd\hat{x}_2,
\end{equation}
where $\mu_i$, $i=1,2$ are dimensionless reduced mass factors corresponding to vectors $\vec{x}_i$.

Upon the hyper-spherical transformation, the spatial part of the three-body wavefunction can be separated into its hyper-radial, hyper-angular, and spherical parts, each of which can be conveniently expanded. The spherical part is written as a coupled product of spherical harmonics $Y_{l_1m_1}(\hat{x}_1)$ and $Y_{l_2m_2}(\hat{x}_2)$. The hyper-angular part is expanded over eigenstates $\varphi_K^{l_1 \, l_2} (\theta)$ of the grand-angular operator appearing in the hyper-radial Schr\"odinger equation. These functions are explicitly defined in terms of Jacobi polynomials $P_{n_{jac}}^{l_1+\frac{1}{2},l_2+\frac{1}{2}}$ of the order $n_{jac}=0,1,2,\ldots$ with hyper-momentum $K = l_1 + l_2 + 2n_{jac}$:
\begin{equation}\label{eq hyper-angular basis}
   \varphi_K^{l_1 \, l_2} (\theta) =
      N_K^{l_1 \, l_2}
      (\mu_1 \mu_2)^{3/4}
      \sin^{l_1} \theta
      \cos^{l_2} \theta
      P_{n_{jac}}^{l_1+\frac{1}{2},l_2+\frac{1}{2}}(\cos 2\theta).
\end{equation}
The normalization constant $N_K^{l_1 \, l_2}$ is chosen to make these functions orthonormal with respect to a weight from the hyper-spherical volume element in \equ{eq hyper-volume}:
\begin{equation}
   \int_0^{\pi/2} \varphi_K^{l_1 \, l_2} (\theta)
                  \varphi_{K'}^{l_1 \, l_2} (\theta)
                  \left( \mu_1\mu_2 \right)^{-3/2}
                  \sin^2 \theta \cos^2 \theta \,
                  \difd \theta = \delta_{K,K'}.
\end{equation}

Having in mind the importance of the asymptotic behavior of the wavefunction in \equ{eq 3body asymptotics}, the three-body model in \cite{thompson04} employs a hyper-radial basis from \cite{erens71} with the desired exponential trend built into it:
\begin{equation}\label{eq hyper-radial basis}
   \mathcal{R}_{n_{lag}} \left( \rho,\rho_0 \right) =
      \frac{1}{\rho_0^3}
      \sqrt{ \frac{n_{lag}!}{\left(n_{lag}+5\right)!} } \,
      L^5_{n_{lag}}\left(\frac{\rho}{\rho_0}\right)
      \exp \left( -\frac{1}{2} \frac{\rho}{\rho_0} \right),
      \qquad n_{lag}=0,1,2,\ldots.
\end{equation}
Here, $L^5_{n_{lag}}$ are associated Laguerre polynomials of the order $n_{lag}$. Note that this basis is just a suitable mathematical basis whose elements cannot be interpreted as hyper-radial eigenfunctions of the physical three-body system. Basis functions $\mathcal{R}_{n_{lag}}$ are orthonormal with respect to the weight factor $\rho^5$ from the volume element in \equ{eq hyper-volume}:
\begin{equation}
   \int_0^{\infty}
   \mathcal{R}_{n_{lag}} (\rho, \rho_0) \,
   \mathcal{R}_{n_{lag}'} (\rho,\rho_0) \,
   \rho^5 \,
   \difd \rho = \delta_{ n_{lag}, n_{lag}' }.
\end{equation}
The completeness of the hyper-radial basis for any $\rho_0>0$ allows one
to use any value of $\rho_0$, and yet reconstruct the asymptotically correct form of \equ{eq 3body asymptotics} determined by an a priori unknown three-body binding energy. The index $n_{lag}$ is independent from quantum numbers attached to spherical and hyper-angular parts of the wavefunction.

Based on these three-body arguments, we have chosen for each valence term in \equ{eq MiCH wavefunction} the following form:
\begin{equation}\label{eq valence term}
   \psi_{\Gamma_{val} \, J_{val}^{\pi} \, 1-1 } =
   \mathcal{R}_{n_{lag}} \left(\rho,\rho_0\right) \,
   \mathscr{Y}_{\gamma_{val} \, J_{val}^{\pi} 1 -1 } (\theta,\hat{x}),
   \qquad
   \Gamma_{val}=\{ n_{lag}, \rho_0, \gamma_{val} \},
\end{equation}
where $\mathscr{Y}$ is a generalized hyper-harmonic function in LS-coupling:
\begin{equation}\label{eq hyper-harmonic}
   \mathscr{Y}_{\gamma_{val} \, J_{val}^{\pi} 1 -1 }
      (\theta,\hat{x}) =
   \varphi_K^{l_1 \, l_2} (\theta)
   \bigg[
      \left[
         Y_{l_1}(\hat{x}_1) \otimes Y_{l_2}(\hat{x}_2)
      \right]_L
      \otimes
      \left[
         \chi_{n_1} \otimes \chi_{n_2}
      \right]_S
   \bigg]_{J_{val}^{\pi}}
   \left[
      \tau_{n_1} \otimes \tau_{n_2}
   \right]_{1-1}.
\end{equation}
Here, $\chi_{n_i}$ and $\tau_{n_i}$ are spinors and isospinors of valence neutrons. The composite index $\gamma_{val}$ comprises all other numbers as well as the Jacobi channel identifier Y or T, i.e. $\gamma_{val}=\{K, l_1, l_2, L, S, \mathrm{Y/T}\}$. By construction, hyper-harmonics $\mathscr{Y}$ corresponding to a given Jacobi basis are orthonormal in all components of $\gamma_{val}$. In the T Jacobi basis, the Pauli principle between valence neutrons is satisfied by requiring \mbox{$l_1+S=\,$even}; in the Y Jacobi basis, the exclusion principle is satisfied upon the action of $\mathcal{A}^{val}$ in \equ{eq MiCH wavefunction}. Valence angular momenta $l_1$, $l_2$, $L$, and $S$ are not to be confused with those for the core in \sect{sect core}.


\subsection{The Hamiltonian}
\label{sect hamiltonian}

In this work, the nuclear Hamiltonian includes kinetic energies $T_i$ of all nucleons and two-body nucleon-nucleon potentials $V_{ij}$:
\begin{equation}
   H = \sum_{i=1}^A T_i + \sum_{1=i < j}^A V_{ij}.
\end{equation}
No correction is needed for the kinetic energy of the total center of mass as the wavefunction in \equ{eq MiCH wavefunction} is expressed in relative Jacobi coordinates.

In microscopic calculations, the choice of the effective nucleon-nucleon interaction is of crucial importance unless realistic forces are used. If a model is to have anything to do with the real physical problem, one must make sure that the inter-nucleon force is appropriate for all subsystems appearing in the model. In \nuc{6}{He}, valence neutrons are mostly in spin-singlet configurations \cite{zhukov93,csoto93}. The spin-singlet di-neutron state is unbound; however, many effective nucleon-nucleon interactions, such as the Volkov force \cite{volkov65}, do not distinguish this state from a bound spin-triplet neutron-proton state, the deuteron.

In this work we use the semi-realistic Minnesota interaction \cite{minnesota77}. This force reproduces the most important low-energy nucleon-nucleon scattering data and therefore it does not bind the di-neutron. The force renormalizes effects of the tensor force into its central component and binds the deuteron by the right amount assuming a proton and a neutron in a relative s-wave. It also gives realistic results for the bulk properties of nuclei in the lowest s-shell. Besides the central component, the potential contains spin, isospin, and spin-isospin exchange terms. The potential contains the mixture parameter $u$ which can be tuned slightly to adjust the interaction strength. When supplemented by a spin-orbit force \cite{minnesota70}, the Minnesota interaction reproduces low-energy $\alpha$-nucleon scattering data. In \cite{minnesota77}, it is advised to employ a short-range spin-orbit force; on that merit we use the spin-orbit parameter set IV from Table 1 in \cite{minnesota70}.

In \sect{sect results}, we mostly show results obtained for the Minnesota plus the spin-orbit interaction (MN-SO). For comparative purposes, some calculations were performed in the absence of the spin-orbit force (MN). To reproduce the two-neutron separation energy 0.97 MeV of \nuc{6}{He}, the mixture parameter $u$ is changed from its default value 1.0 to 1.015 (MN-SO) and 1.15 (MN). By doing so, we expect to obtain a realistic description of the halo even though the absolute binding of \nuc{4}{He} and \nuc{6}{He} is not reproduced. Essentially, the interaction mixture parameter is the only free parameter in our model. The Coulomb interaction is neglected as it should not
affect the long-range behaviour of the neutron halo.


\subsection{Two-nucleon overlap functions}
\label{sect overlap}

Once the full wavefunction in \equ{eq MiCH wavefunction} is optimized as outlined in \sect{sect computation}, one can use it to calculate various observables. Although most of the observables we calculate are standard, two-nucleon overlap functions deserve special attention. For general considerations regarding these functions, see \cite[sect. 16.4.2]{satchler83} and \cite{bang85}.

The overlap integral between a two-neutron halo nucleus ($\Psi$) described microscopically by \equ{eq MiCH wavefunction} and its own core in the ground state ($\Phi$) is defined as:
\begin{equation}\label{eq overlap integral}
   \mathcal{I}_{M_{J_{val}} 1 -1} =
   \sqrt{\binom{A}{2}}
   \left\langle
      \Phi_{J_{core}^{\pi} M_{J_{core}} \, T_{core} \, M_{T_{core}}} |
      \Psi_{J^{\pi} M_J \, T \, M_T}
   \right\rangle.
\end{equation}
The binomial factor accounts for the number of combinations to pick two out of $A$ nucleons. The integration is done over all degrees of freedom in the core, and so the overlap integral $\mathcal{I}$ depends only on the degrees of freedom of two valence\footnote{In the overlap integral, the two distinct neutrons outside the core are called ``valence" They are not to be confused with neutrons in the valence part of the wavefunction in \equ{eq MiCH wavefunction} where valence and core particles become indistinguishable upon the full antisymmetrization.} neutrons remaining outside the core. For neutron-rich two-neutron halo nuclei, the integral has a well defined isospin and its projection, 1 and -1, respectively, but it does not have a good angular momentum. The integral can, however, be expanded in a complete set of hyper-harmonics of angular momentum $J_{val}^{\pi}$ introduced in \equ{eq hyper-harmonic}:
\begin{equation}\label{eq overlap integral expansion}
   \mathcal{I}_{M_{J_{val}} 1 -1} =
   \sum_{\gamma_{val} \, J_{val}^{\pi}}
   C_{J_{core} M_{J_{core}} J_{val} M_{J_{val}}} ^{J M_J}
   \mathcal{O}_{\gamma_{val} \, J_{val}^{\pi}} (\rho)
   \,
   \mathscr{Y}_{\gamma_{val} \, J_{val}^{\pi} 1 -1 },
\end{equation}
where $C$ are Clebsch-Gordan coefficients. The expansion is carried out in the T Jacobi basis where the hyper-harmonics $\mathscr{Y}$ satisfy the Pauli principle between valence neutrons by construction (see \sect{sect valence}). The hyper-radial part in \equ{eq overlap integral expansion} is not expanded in the Laguerre basis from \equ{eq hyper-radial basis} because the basis functions $\mathcal{R}_{n_{lag}}$ do not have physical significance. Instead,
the overlap functions $\mathcal{O}$ are computed directly from:
\begin{equation}\label{eq overlap functions}
   \mathcal{O}_{\gamma_{val} \, J_{val}^{\pi}} \left( \rho' \right) =
   \sqrt{\binom{A}{2}}
   \left\langle
      \left[
         \Phi_{J_{core}^{\pi} \, T_{core} \, M_{T_{core}}} \otimes
         \mathscr{Y}_{\gamma_{val} \, J_{val}^{\pi} 1 -1}
      \right]_{J^{\pi} M_J \, T \, M_T}
      \left|
      \frac{ \delta(\rho - \rho')}
           { \rho^5 }
      \right|
      \Psi_{J^{\pi} M_J \, T \, M_T}
   \right\rangle
\end{equation}
with the integration carried over degrees of freedom of all nucleons. A meaningful calculation of overlap functions $\mathcal{O}$ requires both wavefunctions $\Phi$ and $\Psi$ to be normalized. Using overlap functions, a three-body-like \mbox{core+n+n} component  of the wavefunction $\Psi$ can be written as:
\begin{equation}\label{eq three-body decomposition}
   \Psi_{J^{\pi} M_J \, T \, M_T}^{overlap} =
   \sum\limits_{ \gamma_{val} \, J_{val}^{\pi} }^{}
   \mathcal{O}_{\gamma_{val} \, J_{val}^{\pi} }
   \left[
      \Phi_{J_{core}^{\pi} \, T_{core} \, M_{T_{core}}} \otimes
      \mathscr{Y}_{\gamma_{val} \, J_{val}^{\pi} 1 -1}
   \right]_{J^{\pi} M_J \, T \, M_T}
\end{equation}
in a form analogous to that of the three-body wavefunction \cite{thompson04}. Note, however, that the core $\Phi$ in \equ{eq three-body decomposition} is fully microscopic, whereas the three-body wavefunction would contain an inert macroscopic core. Generally, the expansion in \equ{eq three-body decomposition} would be a part of the fractional-parentage expansion of $\Psi$.

Overlap functions $\mathcal{O}$ satisfy a three-body-like Schr\"odinger equation with a source term \cite{bang85}. Therefore, at least in the asymptotic region, wavefunctions from three-body models and our microscopically founded overlap function $\Psi^{overlap}$ should behave similarly. On this merit, we can compare our results with those from three-body models at the level of wavefunctions rather than integrated observables. An overlap term characterized by $\gamma_{val}$ is referred to as an overlap or a three-body channel.

For practical considerations, it is convenient to introduce modified overlap functions:
\begin{equation}
   u(\rho) = \rho^{5/2} \, \mathcal{O}(\rho) \,
\end{equation}
with a regular behavior near the origin. Again, in the absence of long-range forces in Borromean-like nuclei, overlap functions decay as:
\begin{equation}\label{eq overlap asymptotics}
   \mathcal{O}(\rho) \xrightarrow{\rho \rightarrow \infty}
   \rho^{-5/2} \exp(-\kappa \rho),
   \qquad
   u(\rho) \xrightarrow{\rho \rightarrow \infty}
   \exp(-\kappa \rho)
\end{equation}
just like three-body wavefunctions in \equ{eq 3body asymptotics}. Formally, the decay constant $\kappa$ in \equ{eq overlap asymptotics} is defined as in \equ{eq 3body asymptotics}, the main difference is that for overlap functions the three-body binding energy needs to be computed microscopically. For the Borromean nucleus \nuc{6}{He}:
\begin{equation}
   E_{3body} = E(^6\mathrm{He}) - E(^4\mathrm{He})
\end{equation}
where $E(^6\mathrm{He})$ and $E(^4\mathrm{He})$ are binding energies corresponding to $\Psi$ and $\Phi$, respectively.

Given the orthonormality of hyper-harmonics in \sect{sect valence}, the norms---spectroscopic factors---of overlap channels are given by:
\begin{equation}
   S_{\gamma_{val} \, J_{val}^{\pi} } =
   \int_0^{\infty}
   \mathcal{O}^2_{\gamma_{val} \, J_{val}^{\pi}} (\rho)
   \rho^5 \, \difd \rho =
   \int_0^{\infty} u^2_{\gamma_{val} J_{val}^{\pi} } (\rho) \, \difd \rho.
\end{equation}
In three-body models, spectroscopic factors give the probability of finding the system in a given channel $\gamma_{val}$.


\section{Numerical details}
\label{sect computation}

\begin{figure}[t!]
   \begin{center}
      \includegraphics[width=0.4\textwidth]
                      {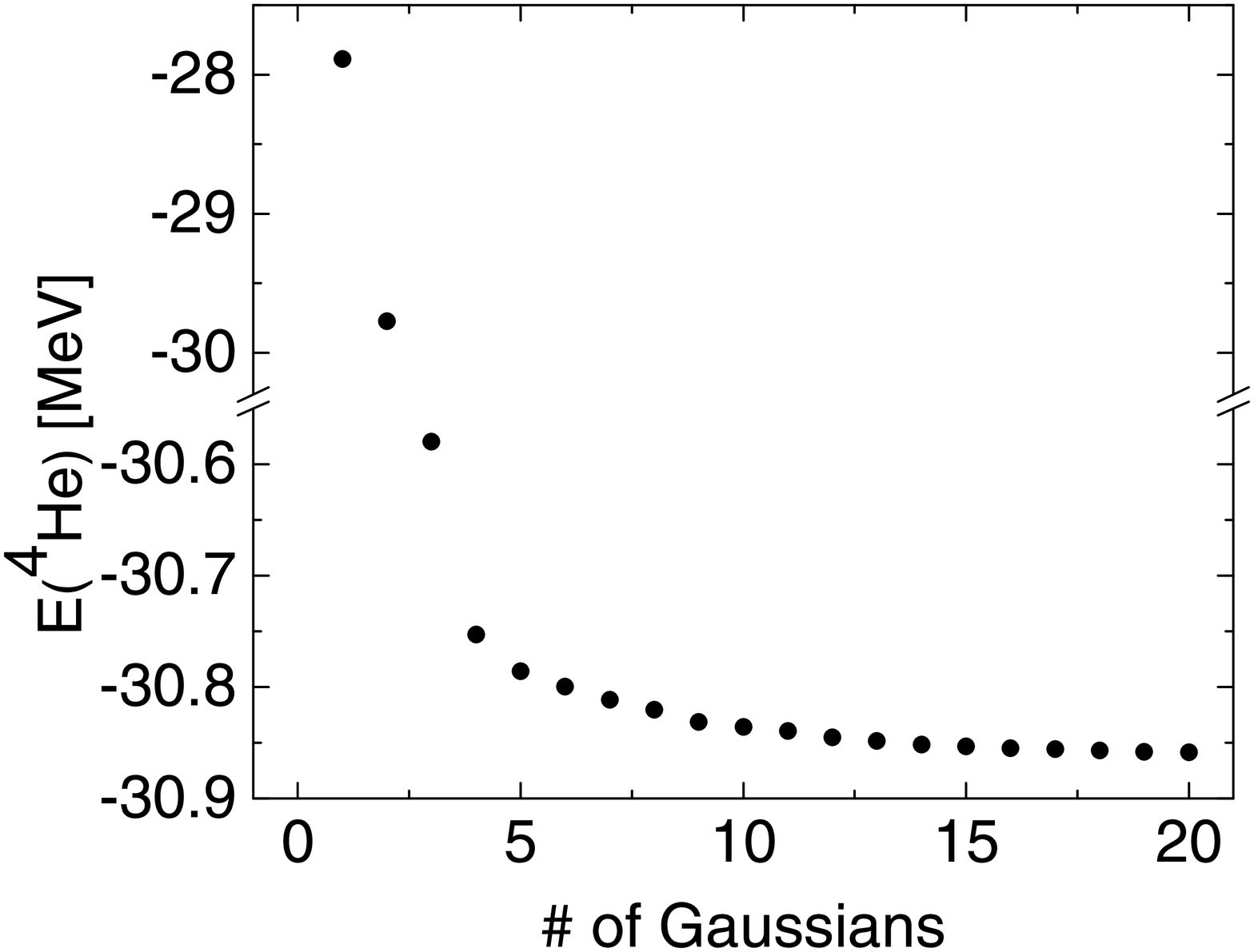}
      \hspace{0.05\textwidth}
      \includegraphics[width=0.4\textwidth]
                      {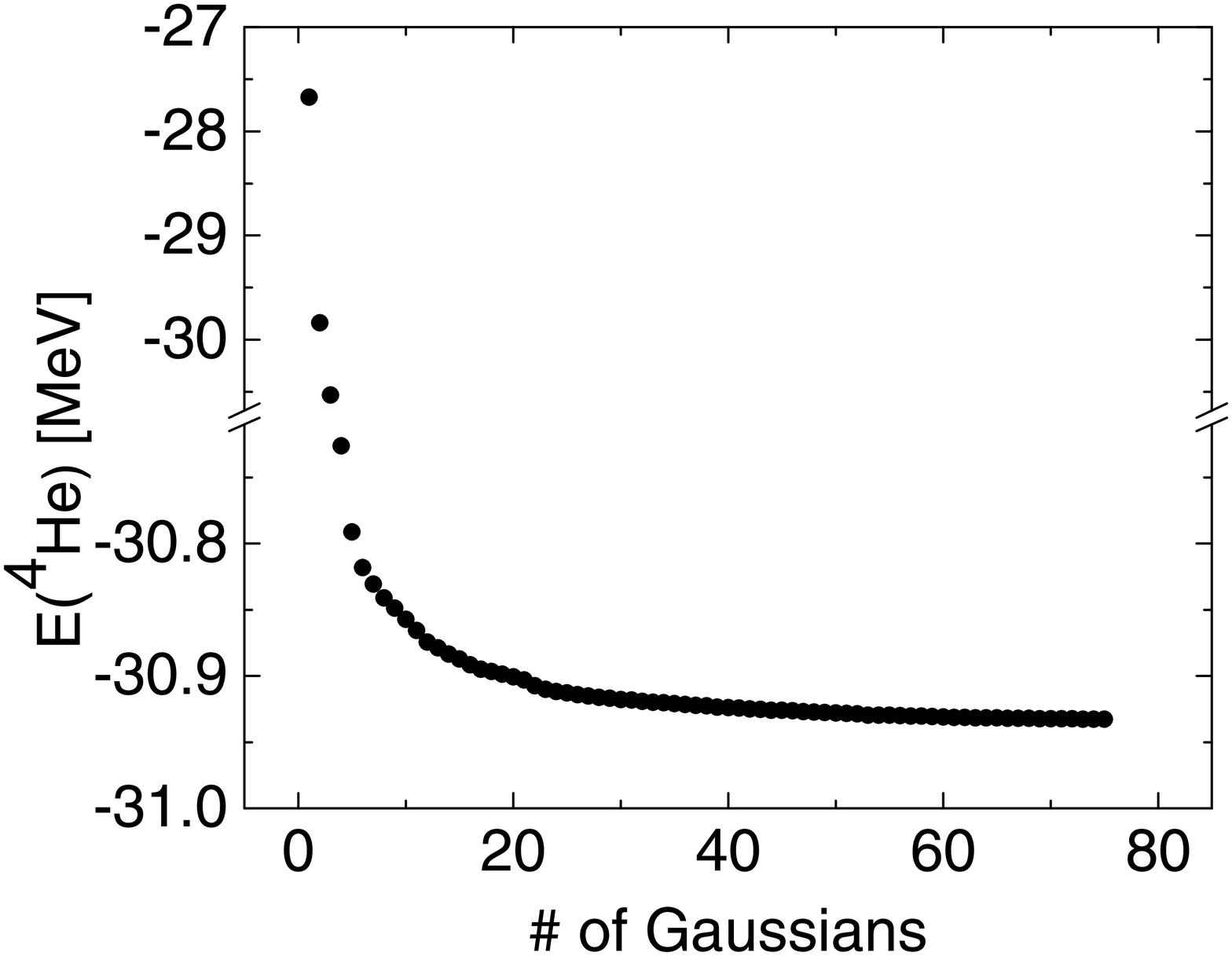}
      \caption{Convergence of the binding energy of \nuc{4}{He}: MN (left) and 
               MN-SO (right).}
      \label{fig 4He energy convergence}
   \end{center}
\end{figure}

In \fig{fig 4He energy convergence}, we show the convergence of the binding energy of \nuc{4}{He} with the number of Gaussians included in the SVM wavefunction described in \sect{sect core}. In converged MN and MN-SO (see \sect{sect hamiltonian}) states containing 20 and 75 basis states, \nuc{4}{He} is bound by  -30.85~MeV and -30.93~MeV, respectively. In both cases, all channels with $l \leq 2$ in \equ{eq core orbital part} were present in the model space. 

The expectation value of energy in \equ{eq expectation value of energy} involves multidimensional integrals which must be evaluated efficiently to perform a meaningful variational calculation. For example in SVM employing the basis of correlated Gaussians, many matrix elements can be evaluated analytically in closed form, and consequently SVM can afford  a random trial-and-error variational search. In our model, however, the core is combined with a functionally very different valence part, and these two become completely entangled upon antisymmetrization. In addition, there does not seem to be an analytical way of computing matrix elements involving different core-valence permuted pieces of the wavefunction in \equ{eq MiCH wavefunction}. For these reasons, we are left with numerical evaluation of all matrix elements. For \nuc{6}{He}, the integrals in \equ{eq expectation value of energy} involve $6 \times 3 = 18$ spatial and $2 \times 6 = 12$ spin-isospin dimensions.

We use techniques of variational Monte Carlo (VMC) \cite{pieper01,vmc-book,foulkes01} to perform the variational search. The integration space is sampled statistically to find the regions most relevant for a given physical problem. In nuclear physics, VMC has been successfully applied to the problem of light nuclei \cite{pieper01}; however, the work here presented is novel in the choice of the trial wavefunction and therefore has its own challenges. In this section, we briefly mention the most important aspects of the optimization procedure, and advise the expert reader to \cite{brida-thesis} for a complete description.

The greatest challenge we faced was to develop a robust and yet efficient method to navigate the parameter space. This aspect falls beyond the framework of VMC. We found that optimization methods used in atomic and molecular physics \cite{lin00,umrigar07} did not work adequately for the problem at hand, most probably due to the specific nature of non-central, state-dependent nuclear interactions. In addition, due to the closeness to the three-body break-up threshold, an arbitrary starting wavefunction is inevitably three-body unbound. As a consequence, since the trial wavefunction drives the scanning of the integration space, all optimization methods tend to break the nucleus into the core and two individual neutrons as long as the system is three-body unbound. The easiest way to solve this pathological problem is to constrain the radius of the nucleus while performing the energy minimization above the three-body threshold. In our model, this is achieved most readily by taking the same non-linear parameters $\rho_0$ in \equ{eq valence term} for all valence terms. Once the system becomes three-body bound,  $\rho_0$, still being the same in all valence terms, is adjusted to minimize the binding energy.

\begin{figure}[t]
   \begin{center}
      \includegraphics[width=0.45\textwidth]
                      {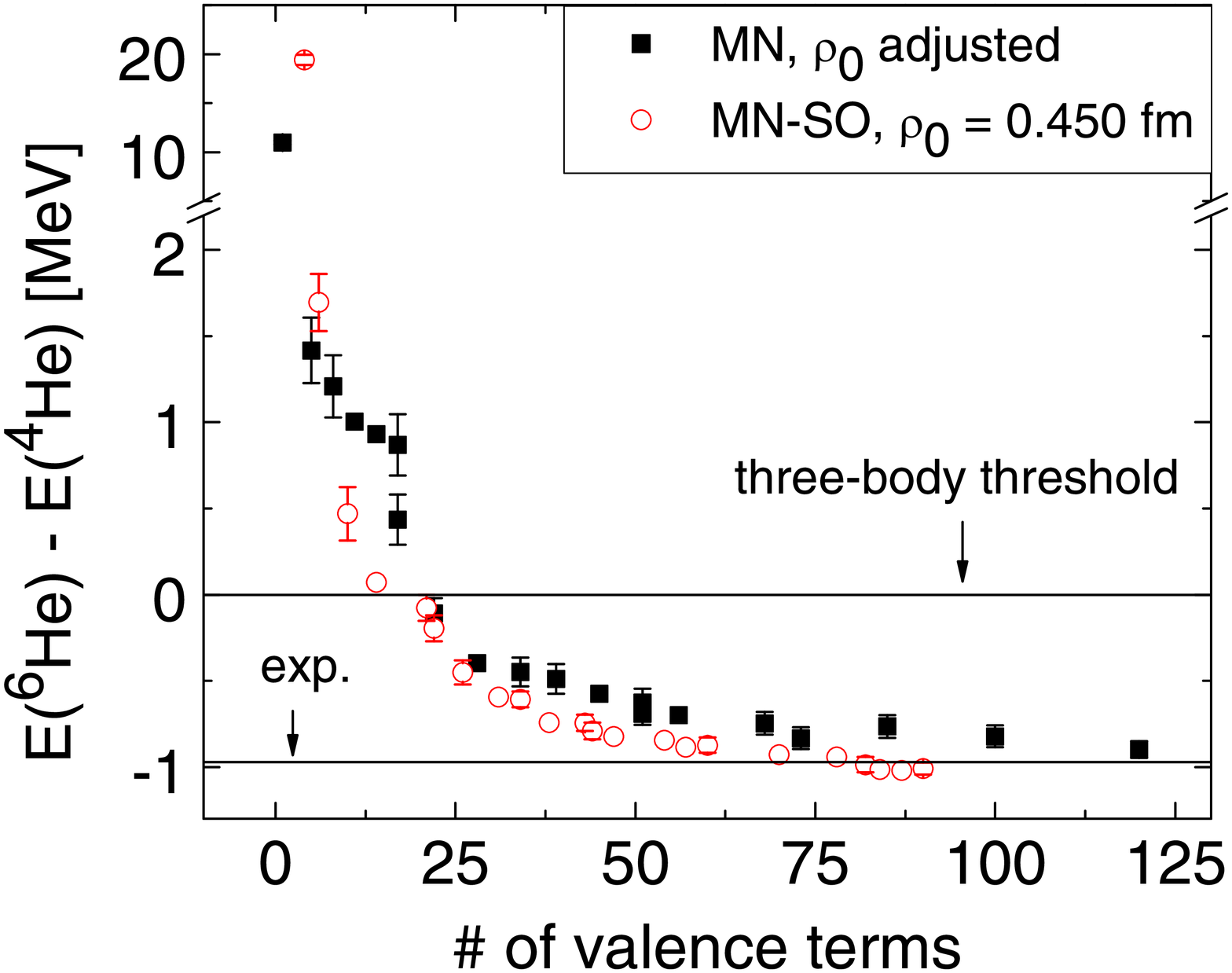}
      \caption{Convergence of the three-body binding energy with the number of
               valence terms included in \equ{eq MiCH wavefunction}. In case MN,
               $\rho_0$ is adjusted along the optimization route; in case
               \mbox{MN-SO}, the results are for fixed $\rho_0=0.45$~fm. Error
               bars were not computed for all points, and even when present they
               may be smaller than actual symbols.}
      \label{fig 6He convergence history}
   \end{center}
\end{figure}

\begin{figure}[t]
   \begin{center}
      \includegraphics[width=0.45\textwidth]
                      {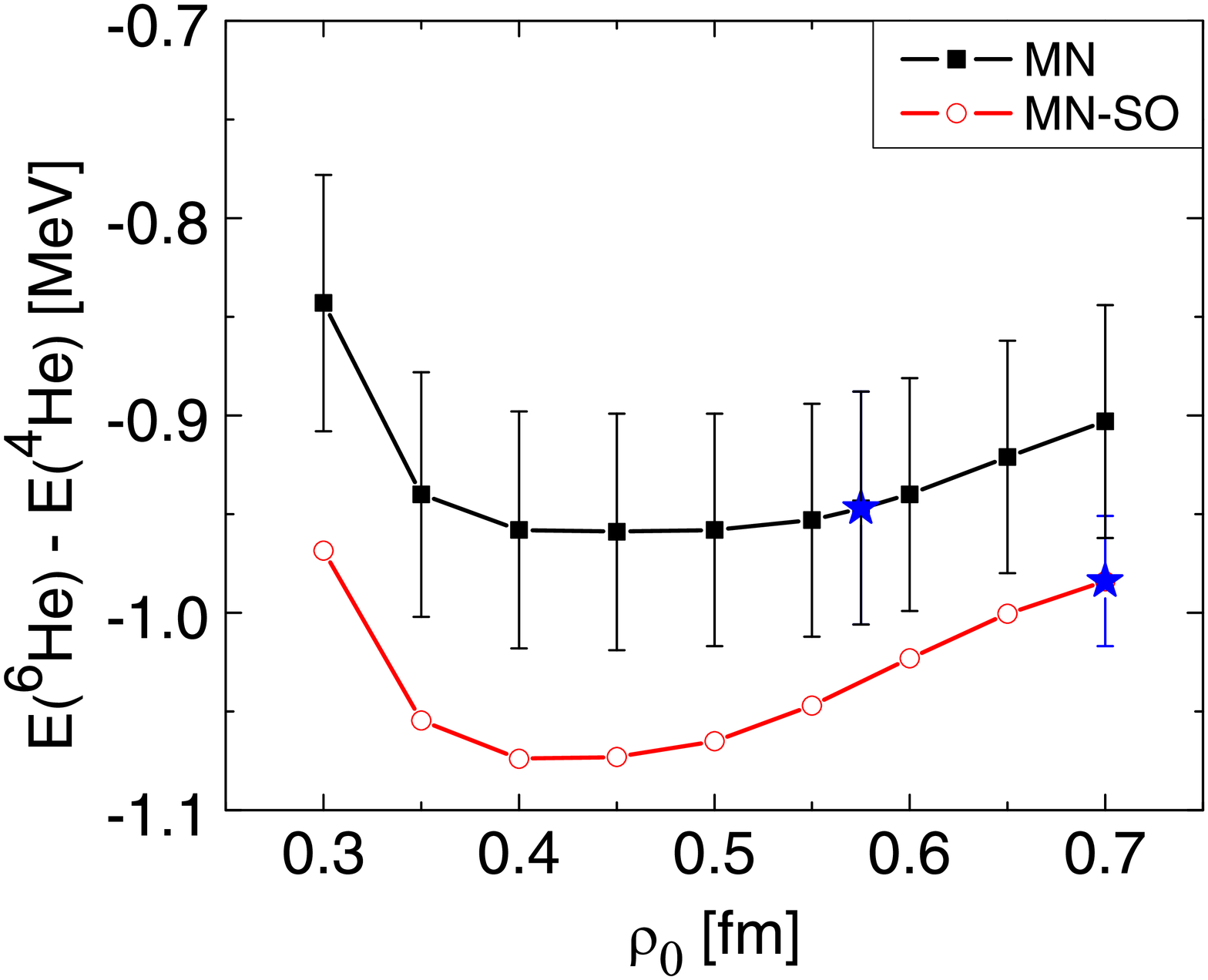}
      \caption{Dependence of the three-body binding energy of converged
               \nuc{6}{He} on the non-linear parameter $\rho_0$. Both curves are                constructed in correlated sampling on walks produced for
               reference values of $\rho_0$ depicted by stars. In the MN-SO case,
               error bars could be computed for the reference $\rho_0$ only.}
      \label{fig 6He minimum search}
   \end{center}
\end{figure}

Different optimization procedures were used for the two interaction cases, MN and \mbox{MN-SO}, of \nuc{6}{He}. In the absence of the spin-orbit force, the MN wavefunction contains only spin-singlet valence terms. It is then possible to simply add hyper-angular and hyper-radial valence terms of increasing orders while re-adjusting $\rho_0$ until convergence in the binding energy is reached, see \fig{fig 6He convergence history}. Starting with a converged wavefunction, different values of $\rho_0$ are tested using correlated sampling to finally locate the energy minimum, as shown in \fig{fig 6He minimum search}. The converged MN wavefunction contains all spin-singlet valence terms with $K \leq 12$ and $n_{lag} \leq 5$ in both Y and T Jacobi bases. For the MN-SO case, the spin-singlet and spin-triplet valence states are mixed, and after antisymmetrization with the core, many components become almost orthogonal which creates numerical noise. It was for this reason that a more advanced optimization technique---comparative optimization on two independent random walks---needed to be developed. The energy minimum with $\rho_0$ is first found for an auxiliary wavefunction as shown in \fig{fig 6He minimum search}, and the MN-SO wavefunction is then tailored to the optimum value of $\rho_0$. Only valence terms lowering the energy the most are admitted to the MN-SO wavefunction at any optimization stage, or in other words $n_{lag}$ and all components of $\gamma_{val}$ in \equ{eq valence term} are treated as discrete variational parameters. It takes ninety carefully selected valence terms with $K \leq 14$ and $n_{lag} \leq 5$ to reach energy convergence in the \mbox{MN-SO} case (see \fig{fig 6He convergence history}). Due to a more selective optimization  process, the MN-SO wavefunction contains fewer valence terms than its MN counterpart.

In both cases, the energy minimum in \fig{fig 6He minimum search} is located at \mbox{$\rho_0 \approx 0.45$~fm}. In either case, the optimization begins with a valence term having low hyper-momentum $K$ and $n_{lag}=0$. Because of their diverging local kinetic energies near the origin, valence terms with $n_{lag} \neq 0$ are avoided until preliminary convergence with $K$ has been reached. To avoid high partial waves in the valence part, both Y and T Jacobi configurations are mixed. The linear coefficients $c$ in \equ{eq MiCH wavefunction} are determined via energy matrix diagonalization. To report final results, values of observables obtained on several independent random walks are averaged for precision. Due to the Monte Carlo integrations, all results come with statistical errors.

\subsection{Three-body calculations}
\label{sect three-body calc}

For comparison purposes, we repeat the inert-core three-body calculations for \nuc{6}{He} published in \cite{danilin98}. Here, we make use of the valence basis introduced in \sect{sect valence}. The phenomenological \mbox{\nuc{4}{He}-n} interaction, vanishing for f- and higher-order partial waves, combines a spherically symmetric central Woods-Saxon and a spin-orbit Woods-Saxon-derivative parts with parameters taken from \cite{danilin98}. As for the interaction between valence neutrons, a realistic Gogny force is used \cite{gpt}. Although the core-n and n-n two-body interactions reproduce $\alpha$-n and n-n phase shifts satisfactorily, \nuc{6}{He} bound by these interactions would miss more than half of its experimental three-body binding energy of -0.97~MeV. As in \cite{danilin98}, the problem of underbinding \cite{zhukov93,funada94} commonly encountered in three-body models based on inert cores is cured with an effective three-body force presumably simulating the effects of the closed \mbox{\nuc{3}{H}+\nuc{3}{H}} channel.

Three-body calculations were performed using the computer code {\sc efadd} \cite{efadd}. The Pauli principle was satisfied by writing the wavefunction in the T Jacobi basis (see \sect{sect valence}) and by projecting the forbidden core-n states before diagonalization \cite{nunes96}. Upon fitting the strength of the ambiguous three-body force, the converged three-body binding energy is -0.98~MeV for the wavefunction containing all valence terms with  \mbox{$K \leq 40$} and \mbox{$n_{lag} \leq 25$}. More three-body results are shown in \sect{sect results} alongside with those from our microscopic model developed in this work.

In three-body models, the size of the core does not enter the actual calculations and some ad hoc assumption about the radius of the core is needed to estimate the size of the whole system. Then, the rms point proton $\langle r_p^2 \rangle^{1/2}$ and rms point nucleon matter $\langle r_m^2 \rangle^{1/2}$ radii of the three-body system (mass number $A$) are related to those of the core (mass number $A_{core}$) through:
\begin{equation}\label{eq three-body radii}
   \langle r_p^2 \rangle^{1/2} =
   \sqrt{
      \langle r_p^2(core) \rangle +
      \langle r^2_{core-CMS} \rangle
   },
   \qquad
   \langle r_m^2 \rangle^{1/2} =
   \sqrt{
      \frac{1}{A}
      \Big[
         A_{core} \langle r_m^2(core) \rangle +
         \langle \rho^2 \rangle
      \Big]
   },
\end{equation}
where $r_{core-CMS}$ is the distance between the core's center of mass and the center of mass of the whole nucleus, and $\langle \rangle$ denote expectation values.


\section{Results}
\label{sect results}

With the fully optimized wavefunctions, we calculate binding energies, rms radii, and density distributions, study core-valence antisymmetrization effects, and extract two-neutron overlap functions for \nuc{6}{He}.


\subsection{Energies and radii}
\label{sect results energies and radii}

Binding energies and radii for \nuc{4}{He} and \nuc{6}{He} from our calculations are shown in \tab{tab results} along with experimental values and results obtained in the three-body model described in \sect{sect three-body calc}, SVM \cite{arai99} representing microscopic cluster models, and ab-initio GFMC calculations \cite{pieper01b}. In \tab{tab results}, experimental rms point proton radii $\langle r_p^2 \rangle^{1/2}$ were computed from accurately measured charge radii $\langle r_c^2 \rangle^{1/2}$ by using the relationship \cite{wang04}:
\begin{equation}\label{eq charge versus proton radius}
   \langle r_p^2 \rangle = \langle r_c^2 \rangle -
                          \langle R_p^2 \rangle -
                          \langle R_n^2 \rangle \frac{N}{Z},
\end{equation}
where $\langle R_p^2 \rangle^{1/2} = 0.895(18)$~fm \cite{sick03} is the rms charge radius of the proton, $\langle R_n^2 \rangle = -0.120(5)$~fm$^2$ \cite{kopecky95,kopecky97} is the mean-square charge radius of the neutron, and $N$ and $Z$ are the neutron and proton numbers, respectively.

Based on arguments in \sect{sect intro} and \sect{sect three-body calc}, there is a qualitative difference between three-body and microscopic results in \tab{tab results}. Strictly speaking, the three-body results should be taken with caution because the three-body binding energy was fitted using an auxiliary three-body force of an arbitrary strength, while for radii we arbitrarily assumed the size of the core in \equ{eq three-body radii} to be equal to that of our MN-SO \nuc{4}{He}. The later assumption was made to ensure the best comparison between three-body and our results. Similar arbitrary assumptions are made for densities from the three-body model in \sect{sect results densities}.

In our model, a free \nuc{4}{He} in \tab{tab results} is overbound and smaller relative to experimental data which sets a wrong scale for absolute binding of \nuc{6}{He}. We are, however, mostly interested in three-body-like features of \nuc{6}{He} which should depend more on three-body rather than the absolute binding energy. As mentioned in \sect{sect hamiltonian}, the mixture parameter in the Minnesota force was adjusted so that the three-body binding energy is about right. Upon this adjustment, there is no additional freedom in computing other observables.

\begin{table}[tb]
\begin{center}
   \caption{Absolute ($E$) and three-body ($E_{3body}$) binding energies in 
            [MeV], and rms point nucleon (proton $p$, neutron $n$, matter 
            $m$) radii in [fm] of \nuc{4}{He} and \nuc{6}{He}. MN and MN-SO are 
            results of this work; for other models, see text. Experimental 
            proton radii were computed from \equ{eq charge versus proton radius}
            using charge radii from references cited in the table. Experimental 
            rms neutron radii were computed from experimental values of proton 
            and matter radii using $\langle r_m^2 \rangle = (1/A)[Z\langle r_p^2
            \rangle + N\langle r_n^2 \rangle ]$. The thickness of the neutron 
            halo is defined as \mbox{$\Delta r = \langle r_n^2 \rangle^{1/2} - 
            \langle r_p^2 \rangle^{1/2}$}.}
   \label{tab results}
   \begin{tabular}{cl r@{.}l r@{.}l r@{.}l r@{.}l r@{.}l r@{.}l}
      \toprule
      & & \multicolumn{2}{c}{MN}    &
          \multicolumn{2}{c}{MN-SO} &
          \multicolumn{2}{c}{3body} &
          \multicolumn{2}{c}{SVM}   &
          \multicolumn{2}{c}{GFMC}  &
          \multicolumn{2}{c}{exp.}
      \\ \midrule
      \multirow{2}*{\nuc{4}{He}} &
      $E$    &
      -30&85 &
      -30&93 &
      \multicolumn{2}{c}{N/A} &
      -25&60   &
      -28&37(3) &
      -28&30 \cite{tilley92}
      \\ 
      & $\langle r_p^2 \rangle^{1/2}$ &
      1&40    &
      1&40    &
      \multicolumn{2}{c}{N/A}    &
      1&41    &
      1&45(0) &
      1&46(1) \cite{sick08}
      \\ \midrule
      \multirow{8}*{\nuc{6}{He}} &
      $E_{3body}$ &
      -0&90(5)    &
      -1&02(3)    &
      -0&98       &
      -0&96       &
      -1&03(10)   &
      -0&97 \cite{tilley02}
      \\ 
      & $\langle r_m^2 \rangle^{1/2}$ &
      2&41(1) &
      2&32(1) &
      2&49 &
      2&42 &
      2&55(1) &
      2&48(3) \cite{tanihata88}
      \\
      & & \multicolumn{2}{c}{} &
          \multicolumn{2}{c}{} &
          \multicolumn{2}{c}{} &
          \multicolumn{2}{c}{} &
          \multicolumn{2}{c}{} &
          2&33(4) \cite{tanihata92}
      \\
      & $\langle r_p^2 \rangle^{1/2}$ &
      1&81(1) &
      1&75(1) &
      1&86 &
      1&81 &
      1&91(1) &
      1&91(2) \cite{wang04}
      \\
      & $\langle r_n^2 \rangle^{1/2}$ &
      2&67(1) &
      2&56(1) &
      2&75 &
      2&68 &
      2&82(1) &
      2&72(4)
      \\
      & & \multicolumn{2}{c}{} &
          \multicolumn{2}{c}{} &
          \multicolumn{2}{c}{} &
          \multicolumn{2}{c}{} &
          \multicolumn{2}{c}{} &
          2&51(6)
      \\ \cmidrule(l){2-14}
      & $\Delta r$ &
      0&86(1) &
      0&81(1) &
      0&89 &
      0&87 &
      0&91(1) &
      0&81(4)
      \\
      & & \multicolumn{2}{c}{} &
          \multicolumn{2}{c}{} &
          \multicolumn{2}{c}{} &
          \multicolumn{2}{c}{} &
          \multicolumn{2}{c}{} &
          0&60(6)
      \\ \bottomrule
   \end{tabular}
\end{center}
\end{table}

Most likely due to the smaller \nuc{4}{He} core, the MN and MN-SO proton radii of \nuc{6}{He} are smaller than they should be. On the other hand, matter radii are comparable with those deduced from experiments. To assess how strongly the radii of \nuc{6}{He} depend on the size of the core, we turn to the three-body model. In a naive three-body picture based on \equ{eq three-body radii}, the larger size of \nuc{6}{He} in the three-body model (when compared to MN and MN-SO) is solely due to the valence neutrons living on average slightly farther from the core. If the radius of the core is increased to its experimental value 1.46~fm, the three-body radii of \nuc{6}{He} become \mbox{$\langle r_m^2 \rangle^{1/2} = 2.51$~fm}, \mbox{$\langle r_p^2 \rangle^{1/2} = 1.90$~fm}, \mbox{$\langle r_n^2 \rangle^{1/2} = 2.77$~fm}, and \mbox{$\Delta r=0.86$~fm}, and the experimental proton radius of \nuc{6}{He} would be seemingly reproduced. A similar shift towards larger radii could be expected for our results if larger cores were involved. Perhaps due to a stronger three-body binding, MN-SO \nuc{6}{He} is slightly smaller than its MN counterpart, but the thickness of the neutron halo remains about the same.

Within SVM, \nuc{6}{He} has been studied in the past repeatedly, e.g. \cite{arai99,varga94}. In \tab{tab results}, SVM results obtained in \cite{arai99} for central and spin-orbit Minnesota and Coulomb interactions are quoted. In that reference, several different cluster compositions were considered to study break-up of the core in \nuc{6}{He}. For \nuc{4}{He}, we
quote the results for the $\alpha_2$ model where the \nuc{4}{He} wavefunction is a superposition of three 0s-harmonic oscillator Slater determinants with common oscillator parameters. Due to this simple picture, SVM \nuc{4}{He} is bound significantly less compared to MN and MN-SO cases. The SVM results for \nuc{6}{He} in \tab{tab results} are those from model (b) in \cite{arai99}. In that model, \nuc{6}{He} was described as a combination of \nuc{4}{He}+n+n and \mbox{\nuc{3}{H}+\nuc{3}{H}} with tritons again built from simple 0s-harmonic oscillators. The triton channel was introduced to overcome the insufficient three-body binding of \nuc{6}{He}. SVM radii and three-body binding energies of \nuc{6}{He} are comparable with ours, especially with the MN model.

For the sake of completeness, we also show ab-initio GFMC results in \tab{tab results}. These were obtained using realistic two-body AV18 and three-body IL2 interactions. The three-body binding energy for this case was computed by using $E(^6\mathrm{He})=-29.4(1)$~MeV \cite{pieper01b}. GFMC results are shown to point out that, by using modern realistic potentials in microscopic calculations, absolute binding energies and proton radii of \nuc{4}{He} and \nuc{6}{He} can indeed be reproduced. However, as we argued in \sect{sect intro} and unless proven otherwise, questions may arise about how well ab-initio models treat asymptotic regions so important for Borromean halo nuclei. Also, two-neutron overlap functions are yet to be extracted from ab-initio models of two-neutron halo nuclei.


\subsection{Density distributions}
\label{sect results densities}

\begin{figure}[t!]
   \begin{center}
      \includegraphics[width=0.5\textwidth]
                      {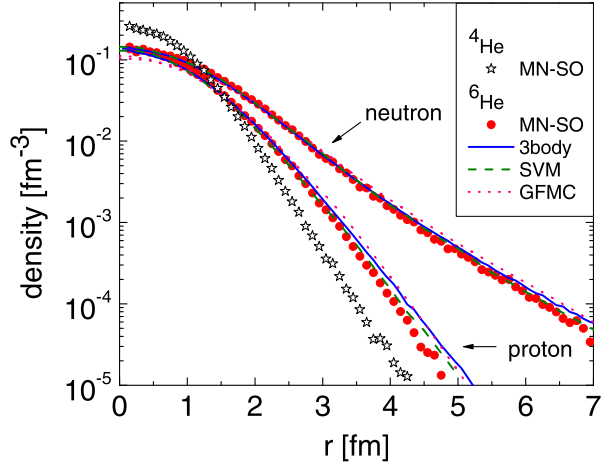}
      \caption{Point proton and neutron densities in \nuc{6}{He} from models in 
               \tab{tab results} (except MN). SVM and GFMC densities are from 
               \cite{arai99} and \cite{pieper01}, respectively; for densities 
               from the three-body model, see text. For comparison, the proton 
               (=neutron) density of the MN-SO \nuc{4}{He} is shown. All proton 
               (neutron) distributions are normalized to the number of protons 
               (neutrons). Here, $r$ is the distance from the center of mass.}
      \label{fig densities}
   \end{center}
\end{figure}

Point nucleon density distributions in \nuc{6}{He} for the more realistic MN-SO case are plotted in \fig{fig densities} along with those obtained from other models listed in \tab{tab results}.
To compute densities from the three-body model, some assumption is needed about the internal structure of the core; quite arbitrarily, we computed these densities using an auxiliary microscopic wavefunction of the type of \equ{eq MiCH wavefunction} constructed by combining the MN-SO \nuc{4}{He} core and the three-body wavefunction serving as the valence part. Once again, this arbitrary construction reveals deficiencies in the few-body models.


The densities in \fig{fig densities} from different models are close to one another with small differences reflecting different radii and wavefunction compositions. All models reproduce the most pronounced property, the neutron halo with the neutron distribution extending far beyond that of protons. Depleted at short distances, the proton density of \nuc{6}{He} stretches farther out than that of \nuc{4}{He}. A partial explanation of this effect comes from the three-body model: in \nuc{6}{He}, the $\alpha$ core does not sit at the center of mass of the entire system, and its motion relative to the center of mass spreads out the proton distribution. Due to the same effect, the neutron density in \nuc{6}{He} is also expected to be depleted at small distances relative to that of a free \nuc{4}{He}, as is also visible in \fig{fig densities}.


\subsection{Antisymmetrization effects}

An important drawback of three-body models is the approximate way in which the Pauli principle between the core and the valence particles is taken into account \cite{thompson00}. The three-body procedure closest to the proper antisymmetrization is Feshbach projection where forbidden states are projected before diagonalization \cite{nunes96}. Given the similarities between the three-body basis and our valence basis, the effects of Pauli blocking can now be assessed microscopically from our model. This can be done simply by including or not including the core-valence antisymmetrizer $\mathcal{A}^{core-val}$ in \equ{eq MiCH wavefunction}. Because of the simpler optimization procedure involved, we performed this study for the MN \nuc{6}{He} only, but qualitatively the same outcome is also expected for the MN-SO case. See \cite{brida-thesis} for details.

The valence channels with $K(=l_1=l_2)=0$ suffer the most from the core-valence Pauli blocking. When $\mathcal{A}^{core-val}$ is not included and the \mbox{$K=0$} valence channels are present in the wavefunction, \nuc{6}{He} is three-body overbound by several tens of MeV. When all $K=0$ channels are removed, the nucleus becomes three-body unbound regardless of the inclusion of valence terms with higher hyper-momenta. When $\mathcal{A}^{core-val}$ is active, converged \nuc{6}{He} including \mbox{$K=0$} valence channels is three-body bound by about -0.9~MeV as shown in \tab{tab results}. Upon removal of all \mbox{$K=0$} channels from the converged wavefunction, the three-body binding reduces to about -0.75~MeV.

It is evident that the proper antisymmetrization is crucial for the structure of \nuc{6}{He}. To produce a meaningful \nuc{6}{He} it is not sufficient to simply neglect the most Pauli-blocked $K=0$ valence channels. Rather, all contributing valence channels ought to be included in the model space and carefully antisymmetrized. Given this conclusion, we strongly advocate using the best available Pauli blocking techniques in three-body models of halo nuclei.


\subsection{Overlap functions}

All models mentioned in \tab{tab results}, although  different in their nature and predictive power, are in fair agreement on the most commonly computed properties of \nuc{6}{He}. To appreciate the amount of microscopic details embedded in different models, it would be better to compare these models at the level of wavefunctions rather than highly integrated observables. In SVM, only s-wave overlap functions for \nuc{6}{He} have been computed, but not expanded in hyper-spherical coordinates \cite{varga94}; in GFMC, these functions are yet to be computed.

Here, we compare overlap functions (\sect{sect overlap}) computed for the MN-SO \nuc{6}{He} to three-body wavefunctions (\sect{sect three-body calc}). In both cases, the \nuc{4}{He} core is in its ground state. The three-body wavefunction is normalized to unity. For a meaningful interpretation of overlap functions, MN-SO wavefunctions of \nuc{4}{He} and \nuc{6}{He} in \equ{eq overlap functions} need to be normalized to unity. For \nuc{4}{He}, the normalization is known analytically from SVM; the norm of the \nuc{6}{He} wavefunction was determined numerically with accuracy of 0.3\% or better by using an auxiliary sampling function \cite{brida-thesis}.

Ordered by spectroscopic factors, the five strongest overlap channels in the MN-SO \nuc{6}{He} are listed in \tab{tab SFs}. These are the only channels that could be resolved, all other potential channels have spectroscopic factors too small and as such are buried in numerical noise. The table also contains three-body results; these were obtained by summing up all hyper-radial components in a given three-body channel. Not only the dominant channels are the same in the two models, but also the order of their spectroscopic strength is preserved. In the three-body model, these five channels account for more than 98\% of the wavefunction. Therefore, we expect that these channels should also grasp most of the \mbox{\nuc{4}{He}+n+n} decomposition of the MN-SO \nuc{6}{He} ground state.

\begin{table}[t]
\begin{center}
   \caption{Spectroscopic factors of the five dominant overlap channels in
            \nuc{6}{He}. All channels are in the T Jacobi basis. Numbers in 
            parentheses are relative errors.}
   \label{tab SFs}
   \vspace{0.1cm}
   \begin{tabular}{ccccccc r@{.}l D{.}{.}{1.2}}
      \toprule
      \multicolumn{6}{c}{channel} &
      \multicolumn{4}{c}{$S$}
      \\
      \cmidrule(l){1-6} \cmidrule(l){7-10}
            &
      $K$   &
      $l_1$ &
      $l_2$ &
      $L$   &
      $S$   &
      3body &
      \multicolumn{2}{c}{MN-SO} &
      \multicolumn{1}{c}{MN-SO / 3body}
      \\
      \midrule
      $K=2$ s-waves &
      2 & 0 & 0 & 0 & 0 & 0.8089 & 1&1155 (0.5\%) & 1.38
      \\
      $K=2$ p-waves &
      2 & 1 & 1 & 1 & 1 & 0.1103 & 0&1859 (0.7\%) & 1.69
      \\
      $K=0$ s-waves &
      0 & 0 & 0 & 0 & 0 & 0.0417 & 0&0555 (2.1\%) & 1.33
      \\
      $K=6$ d-waves &
      6 & 2 & 2 & 0 & 0 & 0.0164 & 0&0266 (3.5\%) & 1.62
      \\
      $K=6$ f-waves &
      6 & 3 & 3 & 1 & 1 & 0.0078 & 0&0122 (3.0\%) & 1.56
      \\ \midrule 
      \multicolumn{6}{r}{$\sum = $} & 0.9851 & 1&3957 &
      \\ \bottomrule
   \end{tabular}
\end{center}
\end{table}

\begin{figure}[t!]
   \begin{center}
      \subfigure[$K=2$ s-waves]{
         \includegraphics[width=0.4\textwidth]
                         {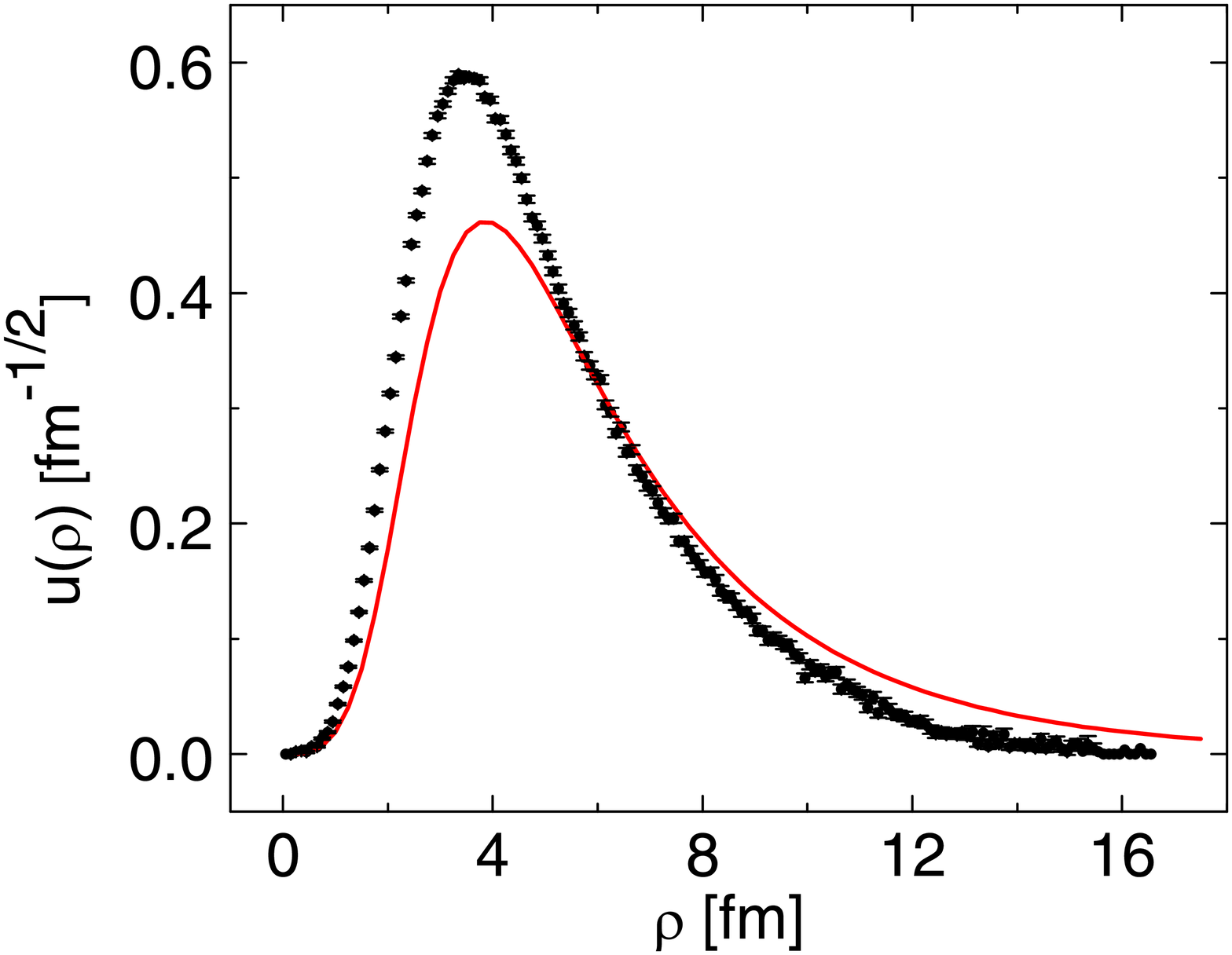}
      }
      \hspace{0.05\textwidth}
      \subfigure[$K=2$ p-waves]{
         \includegraphics[width=0.4\textwidth]
                         {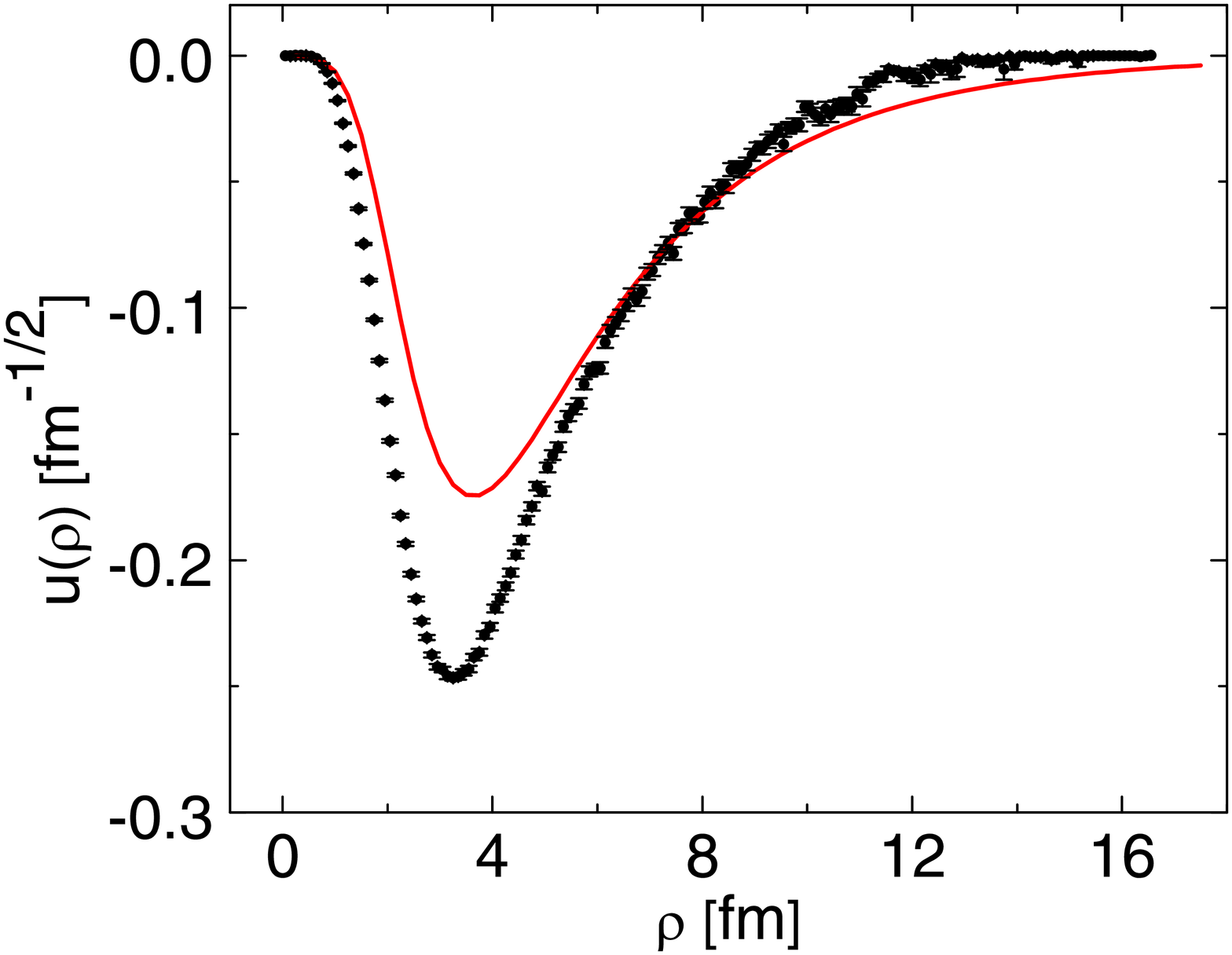}
      }
      \subfigure[$K=0$ s-waves]{
         \includegraphics[width=0.4\textwidth]
                         {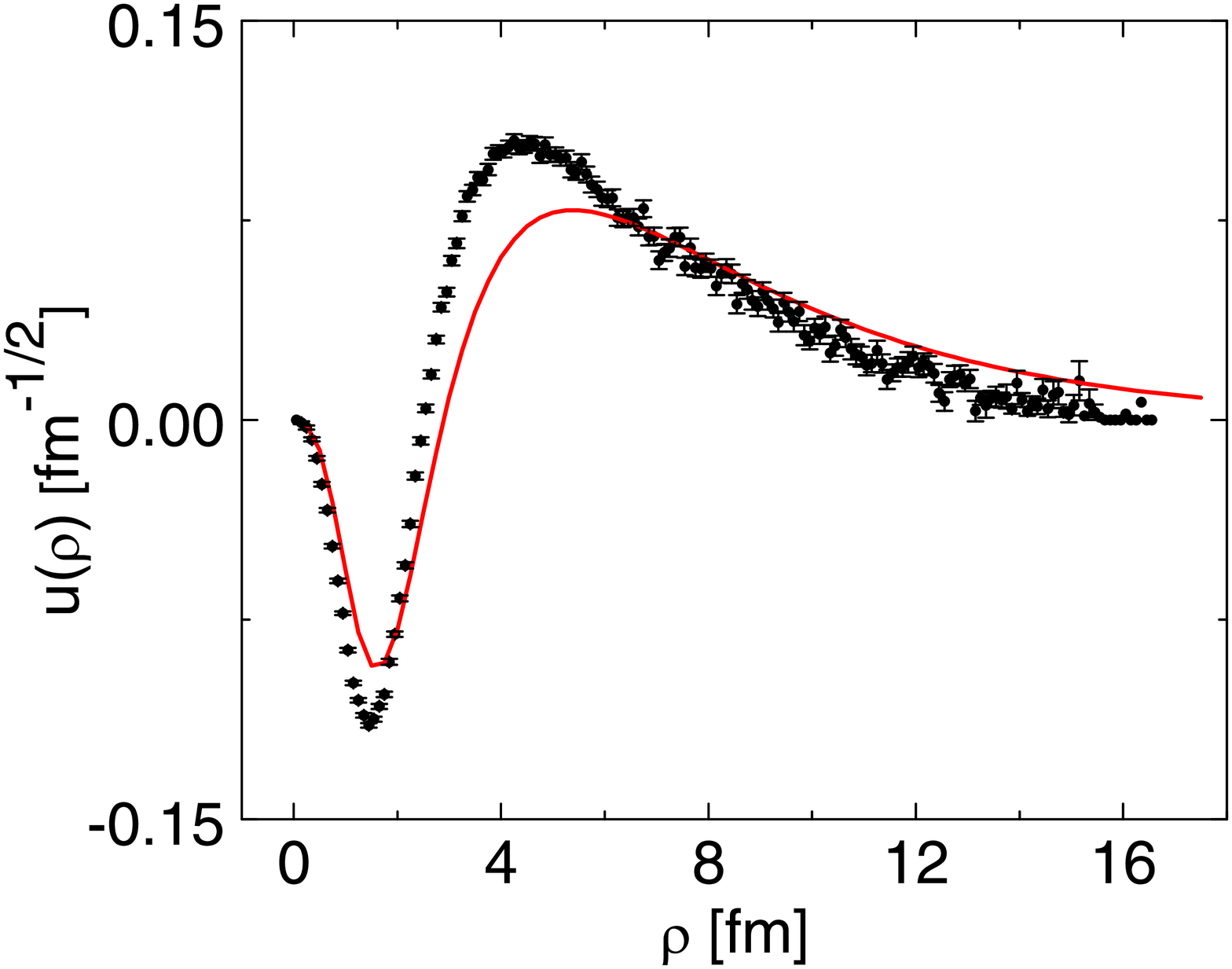}
      }
      \hspace{0.05\textwidth}
      \subfigure[$K=6$ d-waves]{
         \includegraphics[width=0.4\textwidth]
                         {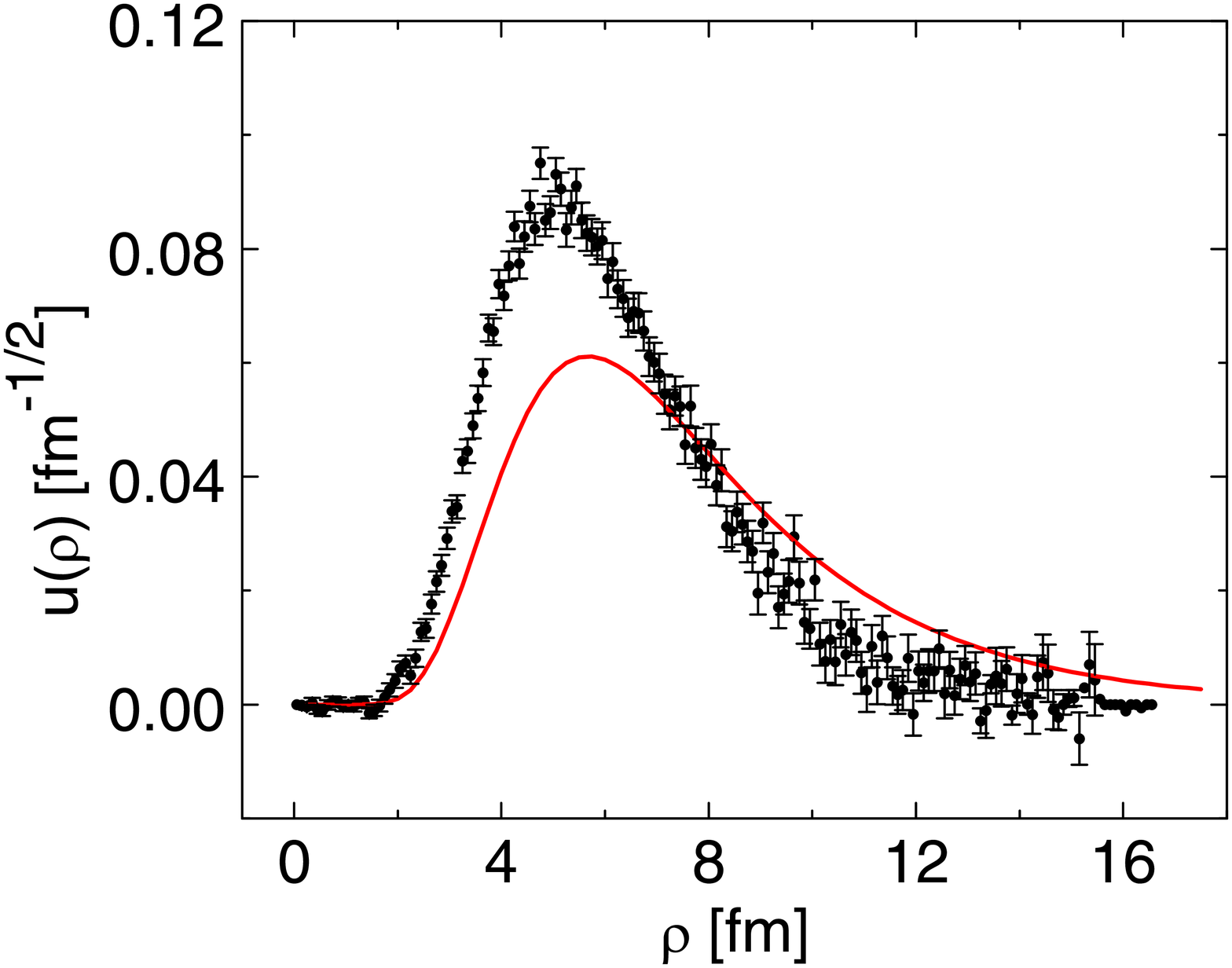}
      }
      \subfigure[$K=6$ f-waves]{
         \includegraphics[width=0.4\textwidth]
                         {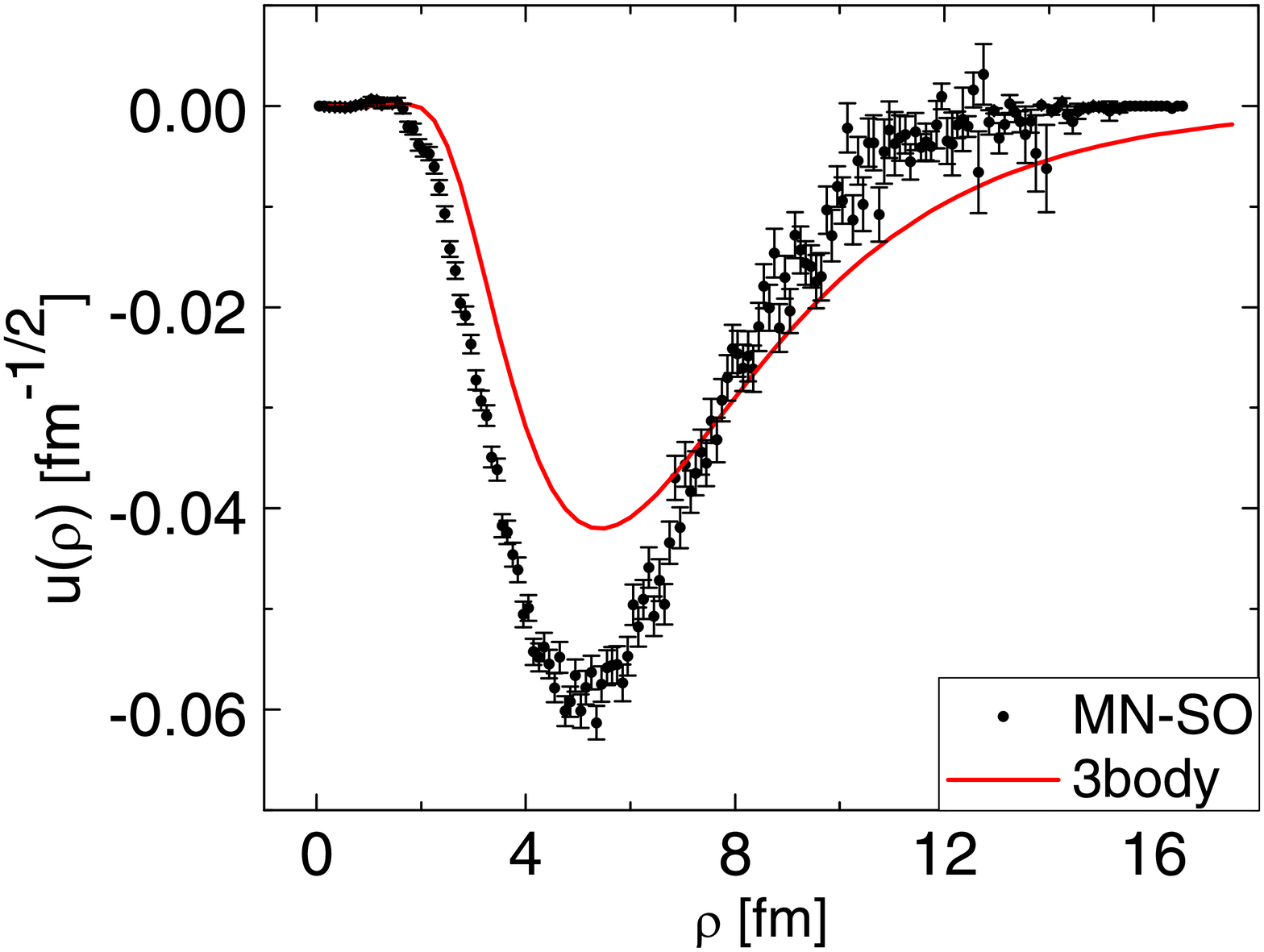}
      }
      \caption{Hyper-radial dependence of overlap MN-SO functions and three-body                wavefunctions for \nuc{6}{He} for those channels presented in \tab{tab SFs}. The
               legend is the same in all panels.}
      \label{fig overlap functions lin scale}
   \end{center}
\end{figure}

\begin{figure}[t!]
   \begin{center}
      \subfigure[$K=2$ s-waves]{
         \includegraphics[width=0.4\textwidth]
                         {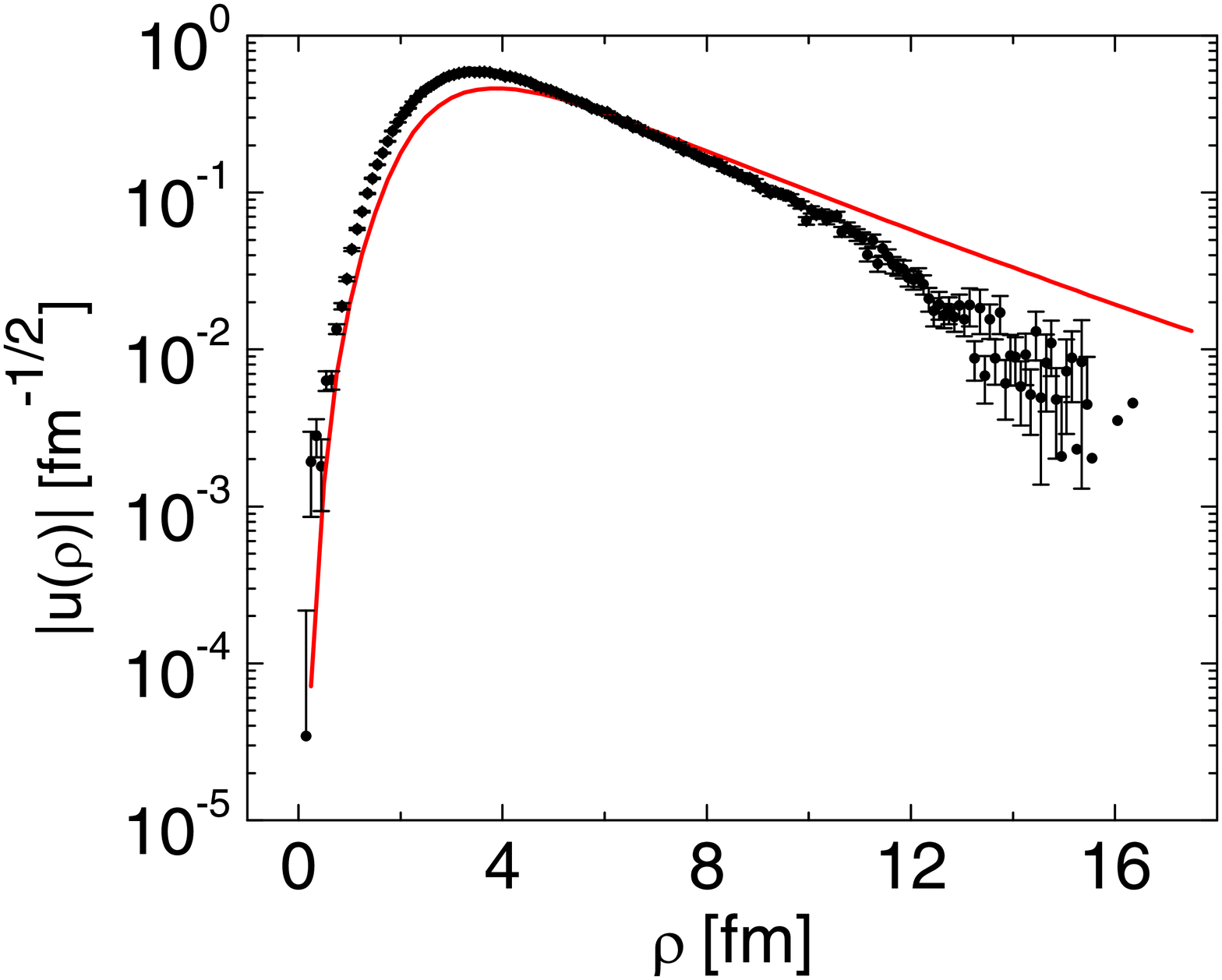}
      }
      \hspace{0.05\textwidth}
      \subfigure[$K=2$ p-waves]{
         \includegraphics[width=0.4\textwidth]
                         {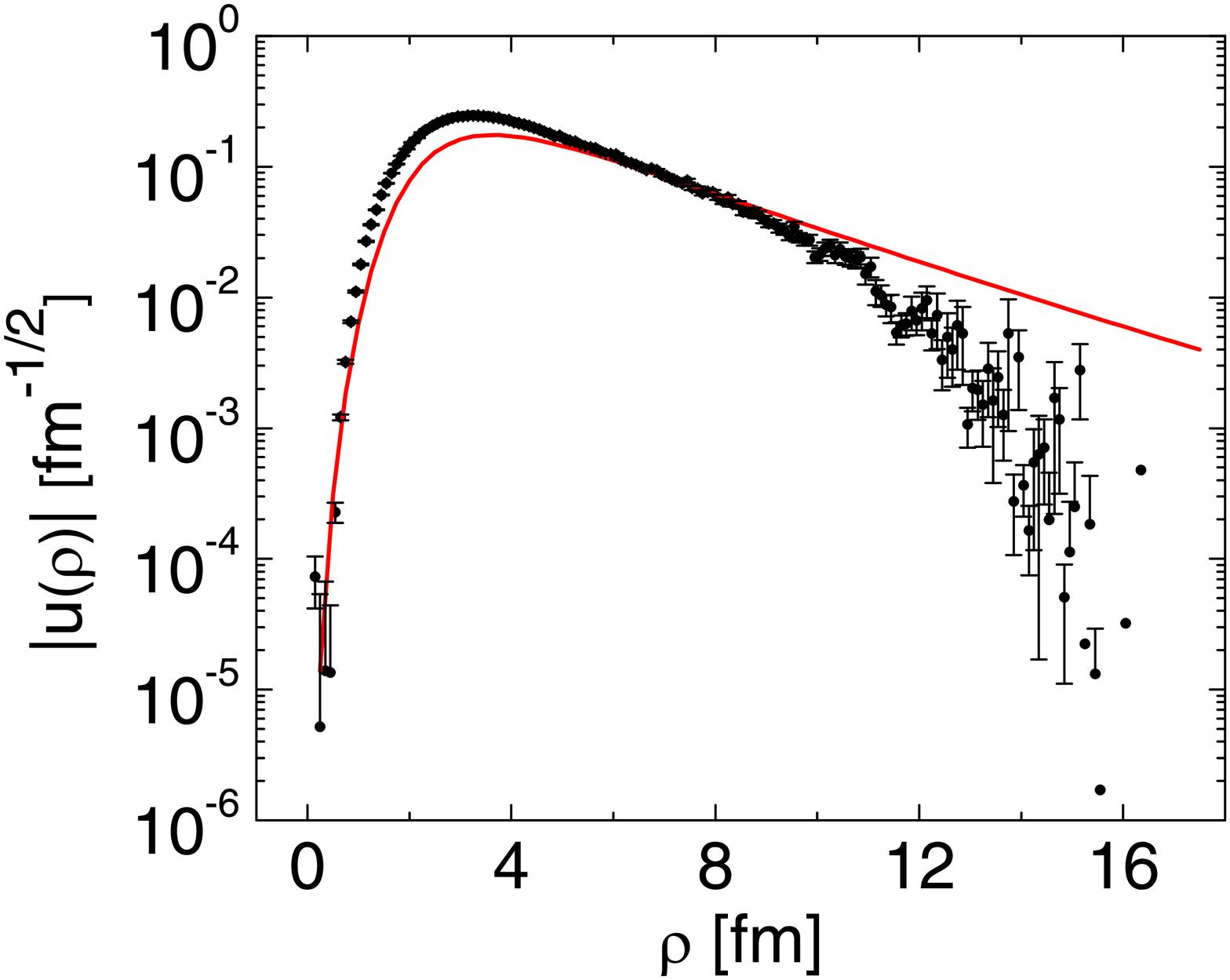}
      }
      \subfigure[$K=0$ s-waves]{
         \includegraphics[width=0.4\textwidth]
                         {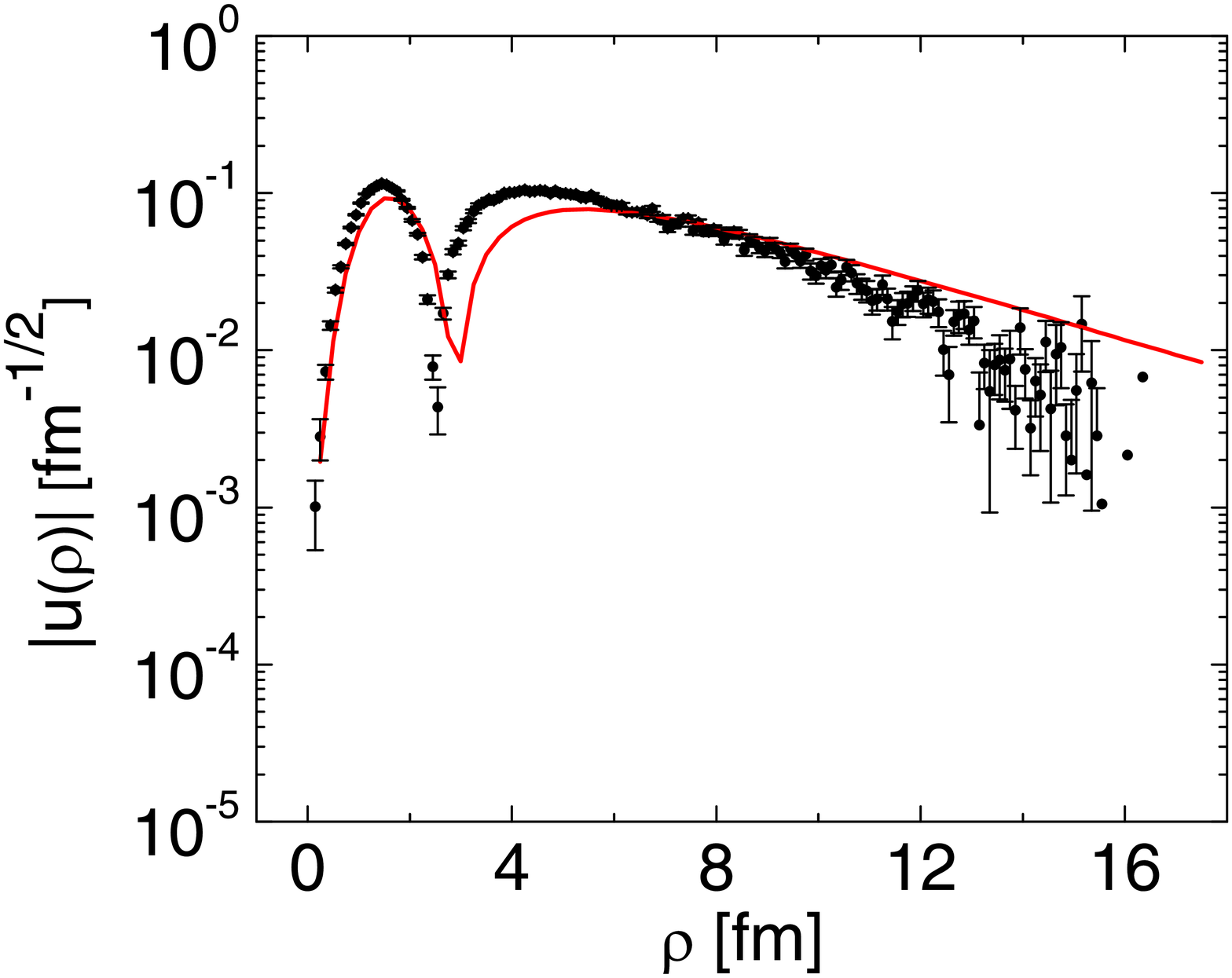}
      }
      \hspace{0.05\textwidth}
      \subfigure[$K=6$ d-waves]{
         \includegraphics[width=0.4\textwidth]
                         {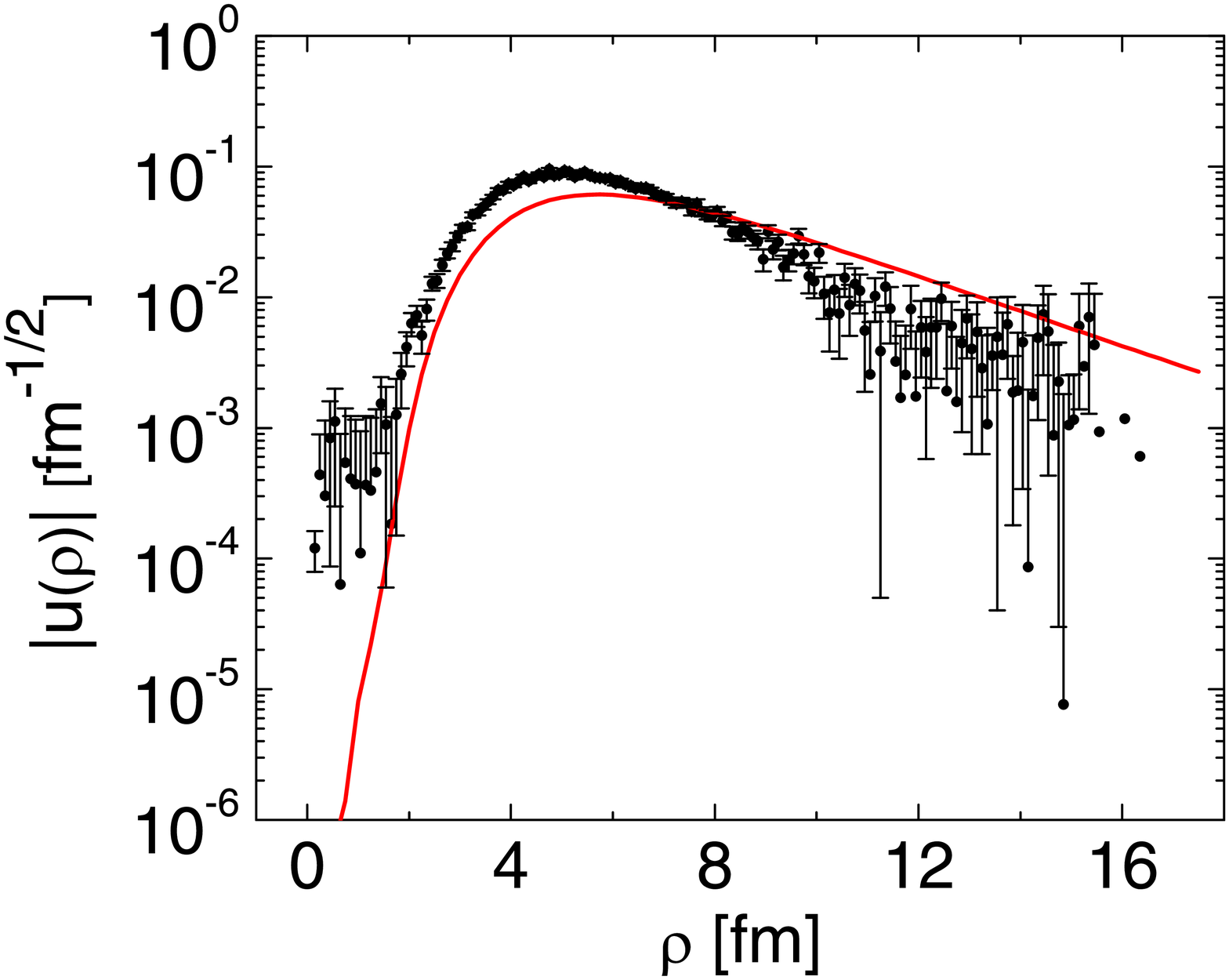}
      }
      \subfigure[$K=6$ f-waves]{
         \includegraphics[width=0.4\textwidth]
                         {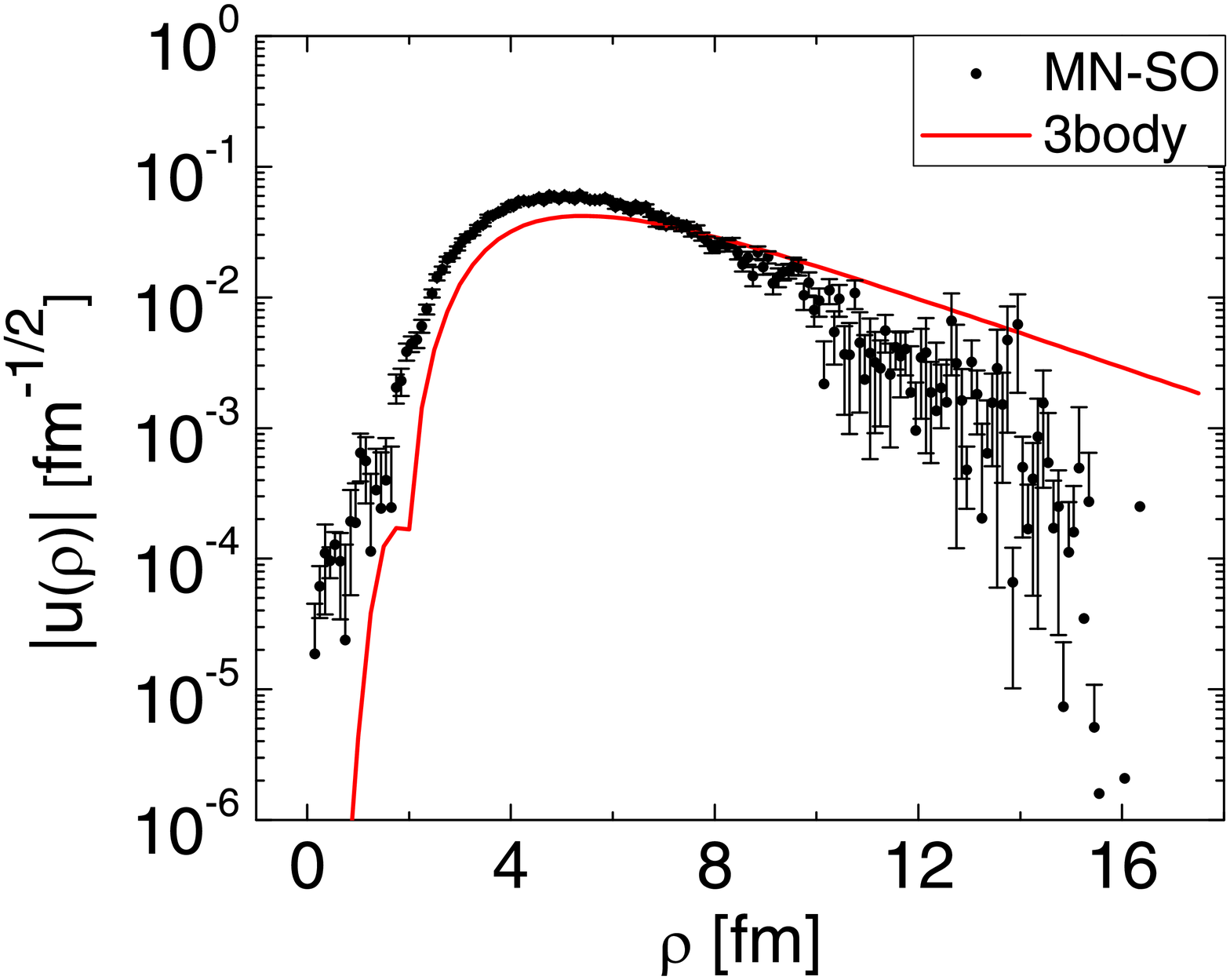}
      }
      \caption{Hyper-radial dependence of absolute values of overlap MN-SO
               functions and three-body wavefunctions for \nuc{6}{He} for those 
	       channels presented in \tab{tab SFs}. The legend is the same in 
               all panels.}
      \label{fig overlap functions log scale}
   \end{center}
\end{figure}

The hyper-radial dependence of overlap channels from \tab{tab SFs} is shown in \fig{fig overlap functions lin scale} and \fig{fig overlap functions log scale}. We show these functions as $u(\rho)$ because of their simpler asymptotic fall-off presented in \equ{eq overlap asymptotics}. The three-body and our model agree on the overall shape of their hyper-radial distributions, and the number of nodes in those channels. On the other hand, our model appears to narrow the hyper-radial distributions and shift their peaks towards smaller distances, and to put more weight on smaller hyper-radii. This difference may be related to smaller radii of the MN-SO \nuc{6}{He} in \tab{tab results} when compared to those of the three-body model.

The most striking difference between our overlap functions and three-body wavefunctions is the normalization of their components quantified in \tab{tab SFs} and visible in \fig{fig overlap functions lin scale} and \fig{fig overlap functions log scale}. The MN-SO spectroscopic factor for the strongest $K=2$ s-waves channel is greater than one due to recoil effects as the \nuc{4}{He} core in \equ{eq overlap integral} does not sit at the center of mass of \nuc{6}{He}. These recoil effects are present in the three-body model too, but there the wavefunction as a whole is normalized to unity and so all spectroscopic factors are smaller than one. This deficiency of three-body models has been pointed out in \cite{timofeyuk01} where an upper limit 25/16=1.5625 was derived for an additional renormalization factor to multiply the three-body spectroscopic factor as a mock-up for missing microscopic information. Indeed, our microscopic model consistently predicts spectroscopic factors greater by at least 30\% than those obtained in the three-body model, but the increase varies between channels in a non-trivial manner as can be seen in \tab{tab SFs}. This observation suggests that to account for microscopic effects, it may not be sufficient to simply renormalize the entire three-body wavefunction by a common factor (such as $\sqrt{25/16}$ suggested in \cite{timofeyuk01}). This conclusion is important, for example, for two-neutron transfer reaction theories for \nuc{6}{He} in which three-body wavefunctions are traditionally used as structure input. 

Since two-neutron transfer cross sections depend on spectroscopic factors, using microscopically derived overlap functions instead of three-body wavefunctions would certainly have implications for reaction observables. For example, if the transfer reaction is mostly sensitive to the dominant component (s-waves) and peripheral, one might not see much change in the normalization, since the asymptotic parts of the overlaps do not significantly change. In the other hand, if the reaction is sensitive to the whole volume of the nucleus, one can expect
a renormalization of the cross section consistent with the spectroscopic factors. Often, two-neutron transfer reactions are complicated by interference of various mechanisms. Preliminary two-nucleon transfer calculations for $^6$He(p,t)$^4$He at $E_{lab}=25$~MeV involving the $^6$He microscopic overlap functions here presented have been performed assuming a 1-step reaction. However, any meaningful comparison with the data requires an extended study of the reaction mechanism which falls beyond the scope of this work.

At hyper-radii beyond about 12~fm, the MN-SO overlap functions in \fig{fig overlap functions lin scale} and \fig{fig overlap functions log scale} become unreliable due to statistical fluctuations. Very large hyper-radii would place valence neutrons into regions very distant from the core, and because the Monte Carlo sampling probability is proportional to the wavefunction squared, such extreme spatial configurations are very unlikely to be visited by a walker during a random walk. Moreover, statistical samples in such distant regions may be highly correlated. For example, in extreme configurations of \nuc{6}{He}, a hyper-radius of 12~fm corresponds to a di-neutron at a distance of 10.4~fm from the center of the core, or to two neutrons on opposite sides of the core, 17~fm apart.

In asymptotic regions, where the two valence neutrons are distant from the core, the core-valence antisymmetrization effects in \equ{eq MiCH wavefunction} disappear. Then, both the three-body wavefunction and the overlap functions should fall off exponentially as shown in \equ{eq 3body asymptotics} and \equ{eq overlap asymptotics}. For $E_{3body} \approx -1$~MeV, we get $\kappa \approx 0.22$~fm$^{-1}$. At larger hyper-radii in \fig{fig overlap functions log scale}, the three-body wavefunctions have the right slope still influenced by the three-body centrifugal barrier. Interestingly, we found in \sect{sect computation} the value $\rho_0 = 0.45$~fm to be optimal for the valence part of the fully antisymmetrized wavefunction of \nuc{6}{He}. By comparing \equ{eq hyper-radial basis} with \equ{eq overlap asymptotics}, the two decay parameters can be related as $\kappa = 1 / (2\rho_0)$. Then,  $\rho_0=0.45$~fm would correspond to $\kappa=1.11$~fm$^{-1}$, or $E_{3body} \approx -25$~MeV. In other words, all individual valence terms in the MN-SO wavefunction decay much faster than the expected asymptotic form of overlap functions. Nevertheless, the strongest MN-SO overlap functions follow approximately the right long-range trend within the computationally safe region. Albeit the efforts in tailoring the basis to the problem, our results still show lingering differences in the asymptotics which could be improved by inclusion of Laguerre polynomials of even higher orders and/or by using a value of  $\rho_0>0.45$~fm in the MN-SO wavefunction.

Even though not discussed in detail, we also computed overlap functions for the MN case of \nuc{6}{He}. Due to the absence of the spin-orbit force, valence spin-triplet states are missing in the MN wavefunction and consequently among overlap functions. Spectroscopic factors of spin-singlet channels were about the same as those for the MN-SO model in \tab{tab SFs}.


\subsection{Clusterization}

\begin{figure}[t]
   \begin{center}
      \includegraphics[width=0.65\textwidth]
                      {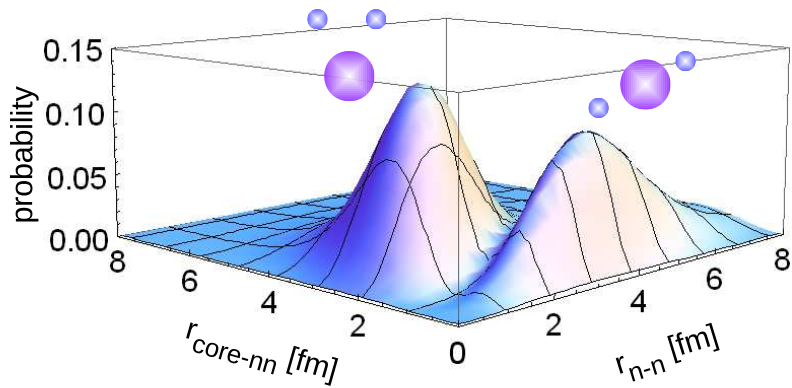}
      \caption{Correlation density plot for the ground state of \nuc{6}{He} 
               using MN-SO. 
      Two dominant clustering patterns are shown
               schematically.}
      \label{fig clustering plot}
   \end{center}
\end{figure}

Although \nuc{6}{He} has been the object of many experimental studies (e.g. \cite{sauvan01,chatterjee08}), there is still an ongoing debate on the neutron clusterization in this nucleus, namely on the dominance of the cigar- or the di-neutron-like configuration. One way of gaining insight into this is by calculating the probability of finding the valence neutrons at definite places within the three-body-like decomposition defined in \equ{eq three-body decomposition}:
\begin{equation}\label{eq clustering probability}
   r_{n-n}'^2 r_{core-nn}'^2
   \frac{1}{2J+1}
   \sum\limits_{M_J}^{}
   \left\langle
      \Psi_{J^{\pi} M_J \, T \, M_T}^{overlap}
      \left|
      \delta(r_{n-n} - r_{n-n}')
      \delta(r_{core-nn} - r_{core-nn}')
      \right|
      \Psi_{J^{\pi} M_J \, T \, M_T}^{overlap}
   \right\rangle,
\end{equation}
where $r_{n-n}$ and $r_{core-nn}$ are the distance between valence neutrons and the distance between centers of masses of the core and the valence neutron pair, respectively. The integration in \equ{eq clustering probability} is carried out over all (unprimed) degrees of freedom of all nucleons. The probability plot for the MN-SO \nuc{6}{He} is presented in \fig{fig clustering plot}. The figure indeed exhibits two peaks, as it should for a system dominated by $K=2$ $s-$waves in the $T-$basis: a di-neutron-like peak positioned at about $r_{n-n}=1.93$~fm and $r_{core-nn}=2.63$~fm ($\rho=3.33$~fm) with the two neutrons close together and far from \nuc{4}{He}, and a cigar-like peak at $r_{n-n}=3.82$~fm and $r_{core-nn}=1.03$~fm ($\rho=2.95$~fm) with the two neutrons positioned on opposite sides of the core. Qualitatively the same clustering picture has been predicted for example within three-body models \cite{zhukov93} and SVM \cite{varga94}. We looked into the clusterization probability distribution produced within the three-body model and found the peaks to be shifted to slightly larger radii (the di-neutron-like peak at $r_{n-n}=2.08$~fm and $r_{core-nn}=2.88$~fm and the cigar-like peak at $r_{n-n}=4.18$~fm and $r_{core-nn}=1.08$~fm) due to the three-body wavefunctions peaking at larger hyper-radii in \fig{fig overlap functions lin scale}. Integrals under the peaks in the microscopic model are larger than those in the three-body model, consistent with the spectroscopic factors presented in Table \ref{tab SFs}: the integral under the di-neutron peak for the microscopic calculation is $0.85$, compared to $0.60$ in the three-body model and the integral under the cigar-like peak for the microscopic calculation is $0.54$, compared to $0.38$ in the three-body model. In both models, it appears that there is coexistence of the two clustering patterns, with $60$\% di-neutron and $40$\% cigar-like configurations.


\section{Summary and Outlook}
\label{sect outlook}

In this work, a microscopic structure model for light two-neutron halo nuclei is developed. Our goal was to combine advantages of few-body and microscopic methods to create a model capable to deal simultaneously with short- and characteristic long-range halo effects exhibited by these nuclei. To succeed, we combine the stochastic variational method and the hyper-spherical harmonic method into a fully antisymmetrized many-body approach. From the computational point of view, our model is similar to the variational Monte Carlo method, the novel feature being the form of the basis which incorporates the few-body features of two-neutron halo nuclei, in particular the correct behavior when the two halo neutrons are far from the core.

The model is applied to the ground state of \nuc{6}{He} bound by an effective Minnesota nucleon-nucleon interaction including the spin-orbit force. When comparing three-body binding energies, radii, and densities, our results are comparable with those from other microscopic models using the same interaction. The overall binding energy is not reproduced as that would require more realistic forces, but this is not essential to produce the characteristic halo features determined by the binding relative to the three-body \mbox{\nuc{4}{He}+n+n} threshold. The halo nature of the nucleus can be seen from its extended neutron density resulting in a large difference between the rms matter and proton radii. We advocate that the standard highly integrated observables, such as binding energies and radii, are not sensitive to details of the halo part of configuration space. To recognize these details, one should compare structure models at the level of wavefunctions or look at observables more sensitive to long-range correlations, such as reaction observables.

We properly calculate, to our knowledge for the first time, microscopic two-neutron overlap functions for \nuc{6}{He} and compare them with those from the state-of-the-art three-body model. These overlap functions provide a crucial input to reaction calculations involving \nuc{6}{He}, in particular to two-neutron transfer reaction models. As our basis was tailored specifically to capture the asymptotic behavior of the wavefunction, it is not surprising that the three-body wavefunction and the overlap functions are indeed similar in this region of coordinate space. In the range of the nuclear interaction, the three-body wavefunction seems to reproduce the properties of the many-body overlap functions only qualitatively. At a quantitative level, however, there are significant differences between overlap functions and three-body wavefunctions. Outstanding amongst these, the spectroscopic factor for the dominant microscopic overlap channel is larger by about 40\% than its three-body counter-part. This discrepancy reveals the deficiency of three-body models, namely their inert-core approximation. Also, our study shows that the underbinding problem in three-body models comes partly from  a poor treatment of antisymmetrization.

It is often thought that few-body models can be corrected for many-body effects by a simple renormalization of the wavefunction. However, we demonstrate that, in general, such a simple fix may not be sufficient because for the most important three-body-like components in \nuc{6}{He}, the ratio between microscopic and three-body spectroscopic factors varies between 1.3 and 1.7 in a nontrivial manner.

In this work we have shown that it is possible to obtain the correct long-range properties of halo nuclei within a fully microscopic approach. This opens up the opportunity to many applications. Studies of two-neutron transfer reactions involving \nuc{6}{He} using our two-neutron overlap functions are underway. Given that in our model all matrix elements are calculated numerically, an extension of the method towards more realistic nuclear interactions should be straightforward. It would also be exciting to apply this method to heavier systems such as \nuc{11}{Li} which would most likely entail more advanced optimization algorithms. On top of that, further code parallelization would be necessary to tackle the factorial growth in dimensionality of the problem.


\vspace{0.5cm}
{\bf Acknowledgments}

We thank Robert Wiringa and Steven Pieper for many discussions on VMC, Kalman Varga for discussions on SVM and for sharing his SVM code, and Ian Thompson for providing details on the previously published three-body results. The calculations were performed at the High Performance Computing Center at Michigan State University. This work was supported by the National Science Foundation through grant PHY-0555893 and the U.S. Department of Energy, Office of Nuclear Physics, under contracts DE-FG52-08NA28552 and DE-AC02-06CH11357.


\end{document}